\documentclass[a4paper,fleqn,usenatbib]{mnras}

\usepackage{graphicx}   % Including figure files
\usepackage{amsmath}    % Advanced maths commands
\usepackage{amssymb}    % Extra maths symbols
\usepackage{subfig}
\usepackage{color}

\usepackage[T1]{fontenc}
\usepackage{ae,aecompl}
\title[The IMF of the solar neighbourhood]{The stellar Initial Mass Function of
the solar neighbourhood revealed by Gaia}
\author[Sollima et al.]{A. Sollima$^{1}$\thanks{E-mail:
antonio.sollima@inaf.it}\\
$^{1}$ INAF Osservatorio di Astrofisica e Scienza dello spazio di Bologna, 
via Gobetti 93/3, 40129 Bologna, Italy\\
}

\date{Accepted 2019 July 26. Received 2019 July 26; in original form 2019 April 11}

\pubyear{2019}

\begin{document}
\label{firstpage}
\pagerange{\pageref{firstpage}--\pageref{lastpage}}
\maketitle

%\label{firstpage}

\begin{abstract}
I use a sample of more than 120,000 stars in the solar neighbourhood with 
parallaxes, magnitudes and colours estimated with unprecedented accuracy 
by the second data release of the Gaia mission to derive the
initial mass function of the Galactic disc. 
A full-forward technique is
used to take into account for the population of
unresolved binaries, the metallicity distribution, the star formation history 
and their variation across the Galactic disk as well as all the observational
effects. The shape of the initial mass function is well represented by a
segmented power-law with two breaks at characteristic masses. It has
a maximum at $M\sim0.15~M_{\odot}$ with significant flattening 
(possibly a depletion) at lower masses and a slope of $\alpha=-1.34\pm0.07$ 
in the range $0.25<M/M_{\odot}<1$. 
Above 1 $M_{\odot}$ the IMF shows an abrupt decline with a slope 
ranging from $\alpha=-2.68\pm 0.09$ to $\alpha=-2.41\pm 0.11$ 
depending on the adopted resolution of the star formation history.
\end{abstract}


\section{Introduction}
\label{intro_sec}

The observational measure of the relative fraction of stars according to their 
stellar masses $\Psi\equiv dN/dM$ is a long-standing challenge in stellar astrophysics. 
In particular, the distribution of stellar masses at birth 
(the so-called "initial mass function"; hereafter IMF) is a key ingredient in
all stellar population synthesis and dynamical simulations of galaxies and
clusters \citep{2006MNRAS.365..759R,2013MNRAS.433.1378L}, it determines the 
stellar mass-to-light ratio \citep{2014RvMP...86...47C} and 
it has a deep relevance in the understanding of the star formation process
\citep{1977ApJ...214..718S}.

In spite of the extensive effort made by many groups, it is still not clear if
the IMF is Universal or whether it depends on some physical parameter \citep[][and 
references therein]{2001MNRAS.322..231K,2010ARA&A..48..339B,2018A&A...620A..39J}.
Among the various theories of star formation developed so far, two main models
provide detailed predictions on the shape and dependence of the IMF on
environmental parameters. 
The first predicts that the IMF is the result of the joint effect of
fragmentation and competitive accretion in a clustered environment
\citep{1978MNRAS.184...69L,2001MNRAS.324..573B,2008ApJ...684..395H}. 
The fragmentation of the molecular
cloud has a mass-dependent efficiency
which scales with the Jeans mass. The value of this critical mass depends on 
the mean molecular weight and on the density of the gas in a complex way
depending on the relative efficiency of several complex processes such as e.g.
cosmic-rays/photoelectric heating, C/O collisional excitation, dust cooling, etc
\citep{1998MNRAS.301..569L,2005A&A...435..611J,2014ApJ...796...75C}. This implies a dependence of the IMF on
the chemical composition and on the original structure of the cloud. 
In the second theory, stars self-regulate their masses balancing the accretion
rate and feedback \citep{1996ApJ...464..256A}. Because of the existence of complex substructures in
molecular clouds, it is not possible to define a single characteristic mass and
the shape of the IMF is determined by the superposition of the stochastic
distributions of many parameters (sound speed, rotation rate, etc.). Also in this
case, both the accretion rate and the stellar feedback have a metallicity
dependence because of their effect on the sound speed and on the
radiation/matter coupling \citep{1996ApJ...468..586A}.   
On the other hand,
observational evidence has provided controversial results so far 
\citep{1989AN....310..127Z,2002Sci...295...82K,2004ApJ...604..579P,
2007A&A...467..117B,2008ApJ...675..163H,2010Natur.468..940V,2012A&A...545A.147H,
2013ApJ...771...29G,2014MNRAS.444.1957D,2017MNRAS.468..319E}.

Both stellar and dynamical evolution modify the IMF at its extremes: massive
stars evolve faster than low-mass ones, so the present-day mass function (PDMF) 
contains only those stars born within a time interval shorter than 
their evolutionary timescales. 
The PDMF is therefore depleted in high-mass stars with respect to the 
IMF by an amount that depends on the star formation history (SFH) of the considered 
stellar system \citep{1998A&A...334..901S}. 
On the other hand, in stellar systems where a significant 
number of long-range interactions occurred, the tendency toward kinetic energy 
equipartition leads low-mass stars to gain orbital energy 
more efficiently than massive ones, possibly reaching the critical energy 
needed to escape \citep{1940MNRAS.100..396S}. The characteristic timescale 
over which a significant exchange of kinetic energy among stars occurs 
is the half-mass relaxation time ($t_{rh}$). 
As time passes and $t_{age}>>t_{rh}$, the mass function (MF) is progressively depleted at
its low-mass end.
So, the MF observed today differs from the IMF.

The most intuitive and straightforward technique to derive the
PDMF is based on the conversion of the 
luminosity (absolute magnitude) distribution of the Main Sequence (MS) stars 
into masses through the comparison with theoretical isochrones \citep{1960ApJ...131..168L}.
This evolutionary sequence does indeed define a locus where stars of different 
masses spend the majority of their lives with luminosities mainly depending on 
their masses \citep{1973A&A....22..121D}.
From an observational point of view, this task is complicated by many factors
introducing significant uncertainties. 

First, the derivation of absolute magnitudes implies the
knowledge of stellar distances. Distances of stellar systems can be derived with
relatively good accuracy using different techniques based on the 
magnitudes of standard candles such as pulsating variables, binaries and stars
at the tip of the Red Giant Branch
\citep{1978ApJ...226..138L,2001ApJ...556..635B,2006MNRAS.372.1675S}. 
The situation is however more complex for
individual stars in the Galactic field which are displaced across two orders of
magnitudes in distance. Until recent years, geometric techniques such as trigonometric
parallaxes were effective in measuring distances only until a few tens of pc from the
Sun with errors as large as 10\% \citep{1997ESASP1200.....E}.
Because the completeness of the observational sample depends on the apparent
magnitude, bright/massive stars are over-represented in a magnitude-limited
sample and a suitable correction accounting for the spatial distribution of stars
with different masses is necessary \citep{1979ApJS...41..513M}. 

Furthermore, the color and magnitude of
MS stars also depend, beside mass, on chemical composition. Therefore, in stellar systems 
containing stars with different
metallicities (such as the Milky Way and galaxies in general) stars with different 
masses overlap across the MS.
The metallicity distribution and star formation rate (SFR) change with the height
above the Galactic disk because of the dependence of the chemical
enrichment efficiency on the density, the long term dynamical evolution 
and the increasing contamination of the thick disc \citep{2019MNRAS.482.2189W}.

Moreover, unresolved multiple systems are observed with luminosities resulting from the sum of
the luminosities of their individual components and therefore have
magnitudes brighter than those of single stars. Many low-mass stars are
secondary components of binary systems and are therefore hidden \citep{1991MNRAS.251..293K}.

Finally, the extinction varies across the sky according to the column density of
the dust interposed between the observer and the stars, making them appear redder and fainter than
they are. Stellar systems at 
large distances cover a small area in the sky and intersect a constant dust
density and are therefore subject to a relatively homogeneous extinction.
Instead, stars
in the Galactic disc are characterized by a variable extinction depending on
their Galactic latitudes and heliocentric distances \citep{2001ApJ...556..181D}.
 
For these reasons, clusters and associations,
formed by thousand of stars with similar ages and chemical composition, lying at the 
same distance and subject to the same extinction, are
ideal benchmarks for this purpose. Unfortunately, 
globular and massive open clusters, the closest clusters for
which the MF can be determined down to the hydrogen-burning limit with a good
statistics and level of completeness, are always older than their typical half-mass
relaxation time. Their PDMFs are therefore not representative of their IMFs
\citep{2012ASSP...29..115M,2017MNRAS.471.3668S}.
Young massive clusters (with ages $<100$ Myr and $M>10^{5}~M_{\odot}$) are all
located in starburst galaxies at distances $>1~Mpc$ \citep{2010ARA&A..48..431P}. 
Because they are compact and distant,
it is possible to resolve only a few bright stars even with the Hubble
Space Telescope (HST) \citep[][and references therein]{2013ApJ...762..123W}.
The measure of the MF in the galaxies of the Local Volume, while requiring a 
careful treatment of their metallicity distributions, has become feasible in
recent years thanks to HST only for the Magellanic Clouds and the closest Ultra 
faint dwarfs. In these galaxies, it has been possible to sample the MF down to a
limiting mass of $\sim0.4~M_{\odot}$
\citep{2006ApJ...641..838G,2013ApJ...763..110K,2018ApJ...855...20G}.
Most of the information on the IMF comes from the associations and star forming
regions. Studies conducted in the nearby associations suggest an average
power-law slope $\alpha\sim-2.3$ \citep{1998ASPC..142..201S,2001MNRAS.322..231K} although 
notable examples of clusters with
flatter \citep[such as the Arches and Quintuplet clusters; 
$-1.9<\alpha<-1.7$;][]{2016MNRAS.460.1854S} or steeper \citep[e.g. NGC 6611;
$\alpha\sim-3$;][]{2014MNRAS.444.1957D} MFs exist. 

The solar neighbourhood is also a privileged site to study the IMF.
Indeed, the IMF modification due to collisional effects in the Galactic disc is negligible and thousand of stars
are observable in the solar vicinity. On the other hand, the lack of accurate 
distances limited the analysis to a small sample of nearby stars.
In his pioneering work, \citet{1955ApJ...121..161S} converted the luminosity
function of the sample available at that time into the MF adopting a
constant star formation rate and neglecting the density/age/metallicity 
variation across the disc and the effect of binaries. This work showed that the IMF is well represented by 
a single power-law $\Psi\propto M^{\alpha}$ with index $\alpha=-2.35$ over the range $0.4<M/M_{\odot}<10$. 
In a following study, \citet{1979ApJS...41..513M} applied the same technique to 
an updated luminosity function, adopting a correction for the spatial
distribution of stars with different spectral types and assuming three simplified
SFHs. They found that the IMF is only weakly dependent on the SFH and provided
different analytical fitting functions. Their IMF steepens at increasing masses 
ranging from $\alpha=-1.4$ at $0.1<M/M_{\odot}<1$ to $\alpha=-2.5$ at
$1<M/M_{\odot}<10$ and $\alpha=-3.3$ at $M>10~M_{\odot}$. 
In a series of papers \citep{1991MNRAS.251..293K,1993MNRAS.262..545K,
1995ApJ...453..358K,2001MNRAS.322..231K,2002Sci...295...82K}
Kroupa and collaborators refined the analysis simulating the effect of a
population of unresolved binaries. By summarizing their
IMF determinations in the Galactic disc and in young clusters they defined a 
broken power-law with index $\alpha=-0.3$ in the substellar regime
($M<0.08~M_{\odot}$), $\alpha=-1.3$ at $0.1<M/M_{\odot}<0.5$ and $\alpha=-2.3$
at $M>0.5~M_{\odot}$.
\citet{2001ApJ...554.1274C,2003ApJ...586L.133C} used a volume-limited sample of stars with known parallaxes 
cleaned from binaries to derive a log-normal IMF which smoothly covers a 
wide range of slopes from $\alpha\sim-1.1$ at $0.1~M_{\odot}$ to
$\alpha\sim-2.9$ at $M=10~M_{\odot}$.
The existence of peaks and breaks in the IMF is relevant in the context of star
formation theories since it determines the existence of characteristic masses
where the efficiencies of some of the physical processes involved might have threshold effects \citep{2006MNRAS.368.1296B}. 
In spite of the different observational samples and functional representations, all 
the quoted studies converged in
defining an IMF slope close to $\alpha=-2.3$ at $M>1~M_{\odot}$ and a flatter
slope at low masses.
In recent years, several works tried to recover the IMF of the solar
neighbourhood using the sample of stars with trigonometric parallaxes provided 
by the Hipparcos mission
\citep{2010MNRAS.404..917D,2015MNRAS.447.3880R} or through a best fit of the colour-magnitude 
diagram obtained from the Tycho-2 mission with synthetic stellar population models 
of the Milky Way \citep{2014A&A...564A.102C,2018A&A...620A..79M}.
All these studies measured MF slopes which all agree
in the sub-solar regime with the previous determinations but are significantly
steeper ($\alpha<-2.8$) at larger masses.
Unfortunately, all the above mentioned studies suffer from either the relatively 
small statistics and the large uncertainties of Hipparcos parallaxes 
or from the uncertainties on the distribution of stars across the disc.

A revolution in the inventory of the solar neighbourhood is provided by the
Gaia astrometric mission \citep{2018A&A...616A...1G}. In particular, 
the 2nd data release of Gaia listed positions,
parallaxes and proper motions
for $\sim1.3\times 10^{9}$ stars across the entire sky with accuracies down to 
$20~\mu as$ and magnitudes and colors with mmag accuracies down to relatively
faint magnitudes ($G<20.7$).

Recently, \citet{2019A&A...624L...1M} used Gaia DR2 magnitudes and parallaxes 
to derive the SFH of the solar neighbourhood. As a byproduct, they constrained the 
IMF using a three-segment power-law with breaks at fixed masses at 0.5 $M_{\odot}$ 
and 1.53 $M_{\odot}$. Their best fit slopes appear to depend on the functional form 
of the SFH, with $\alpha$=-1.4 and -2.5 in the mass 
ranges below/above 0.5 $M_{\odot}$ assuming an exponential SFH, and $\alpha$=-1.3 and -1.9 when 
leaving the SFH free to vary without a pre-imposed parametric shape. While extremely 
valuable, this work was not focussed on the MF determination so that the limited 
flexibility of their MF does not allow to draw conclusions on its detailed shape.

In this paper, I use the Hertzprung-Russel diagram of the solar
neihbourhood provided by Gaia to derive the IMF of this portion of the
Galactic disc. In Sect. \ref{data_sec} the adopted selection criteria applied 
to the global Gaia catalog to define the sample of solar neighbourhood stars
are presented. The algorithms
to determine the IMF in two different mass regimes are described in
Sect.s \ref{lo_sec} and \ref{hi_sec}, respectively. The derived IMF is compared
with those estimated by previous works and with those measured in
other dynamically unrelaxed stellar systems in Sect. \ref{comp_sec}. Finally, I discuss the
obtained results in Sect. \ref{summ_sec}

\section{sample selection}
\label{data_sec}

\begin{figure*}
 \includegraphics[width=15cm]{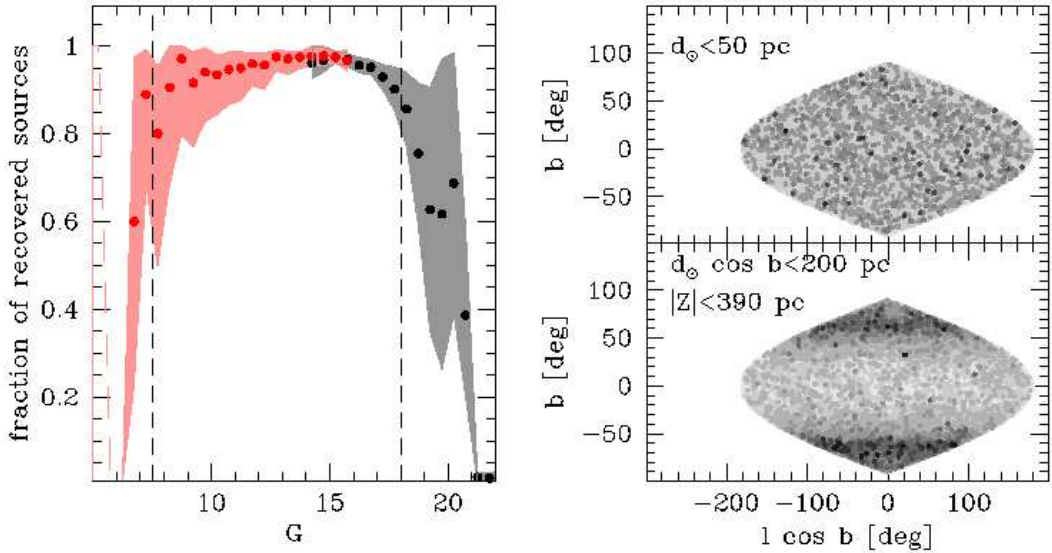}
 \caption{Left panel: fraction of 2MASS (red) and PS1 (black) sources detected
 in the Gaia catalog as a function of the Gaia $G$ magnitude. The shaded areas
 indicate the standard deviation of the 180 analysed fields. Right panels:
 number density of {\it nearby sample} (top) and {\it bright sample} (bottom) stars in
 Galactic coordinates. Darker contours indicate regions with increasing density.}
\label{compl}
\end{figure*}

The analysis presented here is entirely based on data from the second data
release of the Gaia mission \citep{2018A&A...616A...1G}. The Gaia catalog contains
magnitudes in the $G$, $G_{BP}$ and $G_{RP}$ bands, parallaxes ($p$) and 
proper motions ($\mu_{RA}^{*},~\mu_{Dec}$) for $\sim1.3\times10^{9}$ stars in both hemispheres.

Crucial information for the purpose of this work is the catalog completeness.
Stars are included in the Gaia catalog if they are successfully tracked on the
focal plane of the Gaia sky mapper in at least five transits and the resulting
astrometric solution has an astrometric excess noise and semi-major axis of the
position uncertainty ellipse lower than 20 and 100 mas, respectively
\citep{2018A&A...616A...2L}.
According to the Gaia team the catalog is "essentially complete between $12<G<17$" 
while "a fraction of stars brighter than $G<7$ are missing in the data release" with
"an ill-defined faint limiting magnitude" \citep{2018A&A...616A..17A}. Given the complex Gaia selection function, an estimate of the catalog 
completeness can be only made through the comparison with other surveys \citep{2019A&A...621A.144M}.
An attempt in this direction was made by \citet{2018A&A...616A...2L} who derived the fraction 
of stars in the OGLE fields \citep{2008AcA....58...69U} sampled by Gaia DR2 to be $\sim$100\% at G=18 and $>$95\% at G=20 in the Galactic disk.
However, as reported in \citet{2018A&A...616A...2L}, OGLE has a similar limiting magnitude and a 
poorer spatial resolution than Gaia so that such a comparison provides only an 
upper limit to the Gaia completeness.

To derive a conservative estimate, I retrieved the positions and magnitudes from the Gaia archive\footnote{https://gea.esac.esa.int/archive/}
in 180 regions
evenly distributed in the sky covering 3 sq. deg. each and cross-correlated them
with {\it i)} the 2nd data release of the 3$\pi$ Pan-STARRS catalog
\citep[PS1;][]{2018AAS...23143601F}, and 
{\it ii)} the 2Micron All Sky Survey \citep[2MASS;][]{2006AJ....131.1163S}. The 
PS1 catalog covers 3$\pi$ 
steradians and is complete down to $g\sim23.3$ (corresponding to a range in Gaia 
$G$ magnitude between 21.3 and 23.9 depending on the spectral type) i.e. well below the Gaia
limiting magnitude. As most of the catalog samples the Galactic field, crowding is negligible 
and the limiting magnitude is set by the photon noise. 
Therefore, this catalog can be considered suitable to estimate the Gaia completeness at faint
magnitudes. On the other hand, at magnitudes brighter than $G<14$ the PS1 catalog
suffers from photometric saturation, containing many spurious detections
surrounding bright stars. In this magnitude range the 2MASS catalog is
a valid complement since it is based on shallow infrared photometry and is 
complete up to very bright magnitudes. To match the three catalogues, stars within $1\arcsec$ 
were associated and used to construct a colour dependent transformation 
between the PS1 $g$ and 2MASS K magnitudes into Gaia G magnitudes. Stars with a 
magnitude difference $\Delta~G>0.75~mag$ were considered false matches and 
rejected. The fraction of
sources contained in the Gaia catalog as a function of $G$ magnitude is shown
in the left panel of Fig. \ref{compl}. The curves calculated from the comparison with the two 
reference catalogs nicely match in the interval $14<G<16$. The average
fraction of recovered stars is found to be $>90\%$ in the range $7.5<G<18$ with a sudden drop
at fainter magnitudes. At bright magnitudes such a fraction shows large
variations across the sky with a strong dependence on Galactic latitude.
It should be noted that, at low latitudes, most of the incompleteness is caused by 
the high extinction produced by dust clouds in the Galactic disc, whose density
linearly increases with heliocentric distance. This effect is expected to be relatively
low in our analysis, which is restricted to the solar
neighbourhood. So, the curve shown in Fig. \ref{compl} is likely to be a lower limit to
the actual completeness of Gaia. On the basis of these considerations, I selected only stars
in the range $7.5<G<18$ and do not apply any correction to the derived star
counts. 
Of course, the adoption of magnitude cuts creates a bias in the volume-completeness of stars 
with different magnitudes, with the faint (bright) stars being under-represented at large (small) 
distances. This effect must be taken into account in the derivation of the MF 
(see Sect. \ref{metlo_sec} and \ref{methi_sec}).
 
I retrieved from the Gaia archive all stars with measured magnitudes
and parallaxes and applied a quality cut based on the parameters
{\sc astrometric\_chi2\_al} ($\chi^{2}$)
and {\sc astrometric\_n\_goof\_obs\_al} ($\nu$)

\begin{equation*}
\sqrt{\frac{\chi^{2}}{\nu-5}}<1.4~u_{0}(G,G_{BP}-G_{RP})
\end{equation*}

Where $u_{0}$ is a function of colour and magnitude \citep{2018A&A...616A...2L}.
The formal uncertainty on the parallax ($\epsilon_{p}'$) was also corrected using the relation

\begin{equation*}
\epsilon_{p}=\sqrt{(1.08~\epsilon_{p}')^{2}+\epsilon_{p,0}^{2}(G)}
\end{equation*}

with 

\[ \epsilon_{p,0} =
  \begin{cases}
    0.021~mas       & \quad \text{if } G<13\\
    0.043~mas  & \quad \text{if } G>13 \text{~~\citep{2018A&A...616A...2L}}
  \end{cases}
\]

This selection was proven to be effective in removing astrometric
artifacts while preserving the completeness of real stars \citep{2018A&A...616A..17A}.

Gaia is able to resolve binaries with angular separations $>0.4\arcsec$ \citep{2018A&A...616A..17A}, 
corresponding to a physical separation $>30~AU$ at 50 pc. 
This means that most of the binaries in our sample are unresolved.

Absolute $M_{G}$ magnitudes were computed according to the relation
\begin{equation}
M_{G}=G-10+5~\log{\frac{p}{mas}}-E(B-V)~k_{G}
\label{abs_eq}
\end{equation}

The reddening distribution was assumed to follow the relation

\begin{equation}
E(B-V)=\frac{0.03}{sin~b}~\left[Erf\left(\frac{p^{-1} sin~b_{i}+Z_{\odot}}{\sqrt{2}
\sigma_{dust}}\right) - Erf\left(\frac{Z_{\odot}}{\sqrt{2}
\sigma_{dust}}\right)\right]
\label{redd_eq}
\end{equation}

with $\sigma_{dust}=150~pc$.
Such a relation assumes a gaussian distribution of the dust\footnote{The choice 
of a gaussian distribution instead of e.g. a $sech^{2}$ law 
\citep[e.g. ][]{2017MNRAS.470.1360B} is made to allow to derive the dust column 
density using an analytical integration. Note that within the small range covered by the considered 
samples, the maximum difference between the two functional forms is 
$\Delta E(B-V)<0.0015$, much smaller than the typical colour uncertainties.} and has been 
calibrated using the distance along the reddening
vector \citep[from][]{2018MNRAS.479L.102C} in the $(G_{BP}-G)-(G-G_{RP})$ colour-colour diagram of stars in the
{\it bright sample} as a function of their position in the $p-sin~b$ diagram.
The above relation, while neglecting small scale variations within the considered volume, 
well interpolates the reddening map recently obtained by \citet{2019A&A...625A.135L} ($\Delta E(B-V)=0.003 \pm 0.017$).

Different samples were selected to derive the MF in different mass regimes,
which require different treatments:
\begin{itemize}
\item{A sample of nearby ($p>20~mas$) MS stars with $M_{G}>4$ to study the MF in
the low-mass ($M<1~M_{\odot}$) regime (hereafter referred as {\it nearby sample}). 
In this distance range, a star at the hydrogen-burning limit with a solar 
metallicity (with $M_{G}\sim 15.5$) has an apparent magnitude $G<18$, so this sample contains almost all
the low-mass stars contained in this distance range;}
\item{A sample of bright ($M_{G}<4$) stars in a cylinder centred on the Sun
with a radius $R\equiv cos(b)/p<200~pc$ and height above the Galactic plane $|Z|\equiv|sin(b)/p+Z_{\odot}|<390~pc$, to study the
MF in the high-mass ($M>1~M_{\odot}$) regime (hereafter referred as {\it
bright sample}).}
\end{itemize}

The R and Z limits of the {\it bright} sample were specifically chosen to 
avoid nearby dust clouds \citep{2019A&A...625A.135L}. In particular, all the stars within the 
selected volume nicely lie along the relation defined in eq. \ref{redd_eq} reaching a 
maximum extinction of $E(B-V)<0.035$.

I do not apply any correction for the systematic parallax offset reported by \citet{2018A&A...616A...2L}.
These authors quantify this systematic shift to be $\sim$-0.05 mas at 
parallaxes $<$2.8 mas and not clearly measurable at large values, with local 
regional variations depending on colour and magnitude. According to 
\citet{2018A&A...616A..17A} "in absence of a large number of calibrators 
homogeneously spread across the sky in this parallax 
range, it is not advisable to correct it using a simple shift".
Note that, even in the worst case, 
such an offset would be negligible ($<$2\%) in the {\it nearby} sample, while it would be 
partly absorbed by the calibration of the vertical distribution of stars in the 
{\it bright} sample (see Sect. \ref{hi_sec}).

The value of the solar height above the Galactic plane was set as the mode
of the Z distribution of {\it nearby sample} stars at $Z_{\odot}=1.4\pm 0.1~pc$,
which falls between the estimates by \citet[][$-0.9\pm 0.9$ pc]{2017MNRAS.470.1360B} and 
\citet[][$17\pm 5$ pc]{2017MNRAS.465..472K}. Given the vertical extent of our 
sample, the value of $Z_{\odot}$ is relevant only for the {\it nearby} sample. On the 
other hand, small variations in $Z_{\odot}$ do not produce sizeable effects in the analysis.

To avoid domination by the contribution of individual star clusters located
inside the analysed regions, I excluded from the sample the stars belonging to 
two nearby open clusters (Hyades, Pleiades) and those of two other extended coherent
groups not previously identified (at (RA,Dec,p)=(185, -56, 9.2 mas) and
(RA,Dec,p)=(-23, 245, 7.3 mas)). These clusters were
identified on the basis of their clustering in the 5D space of positions, proper
motions and parallax: a star is considered a cluster member if the local 
density is 3$\sigma$ above the background defined by the stars in the 
surrounding portion of this space. Consider that, in the generally accepted scenario, the large 
majority (maybe all) of the field stars form in clusters and associations quickly dissolved after a few Myr.
Therefore, regardless of their clustering in phase-space, all the stars of the solar 
neighbourhood were likely part of a stellar complex at their birth. The selection made here is 
intended to exclude only the contribution of the largest clusters.
Note that these clusters contain only a 
small fraction of the stars in the considered samples ($\sim$1\% of the {\it nearby} sample and 
$<$0.1\% of the {\it bright} sample), so that the results do not depend on this 
choice. 

After applying the above defined selection criteria, the {\it nearby} and 
the {\it bright} samples contain 27048 and 120724 stars, respectively, 
The number density of the stars belonging to the two samples at various positions in the sky are
also shown in the right panels of Fig. \ref{compl}. Stars nicely follow the expected distribution for 
the considered sample volumes with no
apparent patches, excluding a significant position-dependent sampling efficiency
due e.g. to the Gaia scanning law.

In the next two sections I will describe the methods adopted to derive the IMF in the
two above defined mass regimes.

\section{Low-mass regime}
\label{lo_sec}

\subsection{Method}
\label{metlo_sec}

\begin{figure}
 \includegraphics[width=8.6cm]{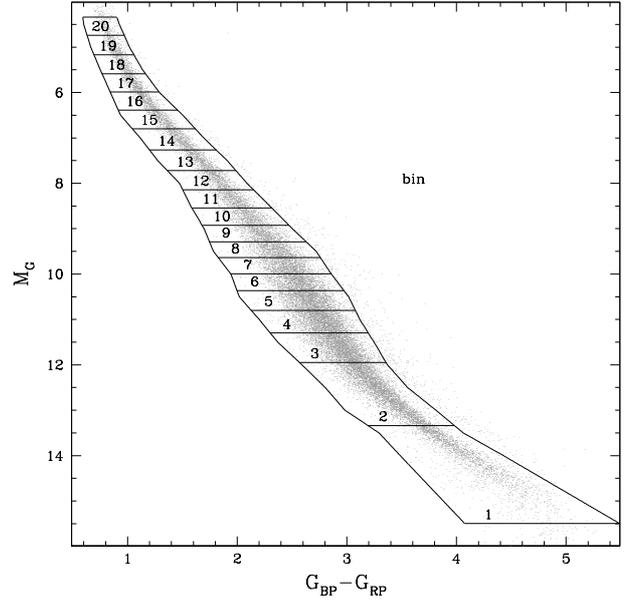}
 \caption{CMD of the {\it nearby sample}. The adopted selection boxes 
 corresponding to the 20 mass-bins (see the text) are shown.}
\label{cmd}
\end{figure}

\begin{figure*}
 \includegraphics[width=15cm]{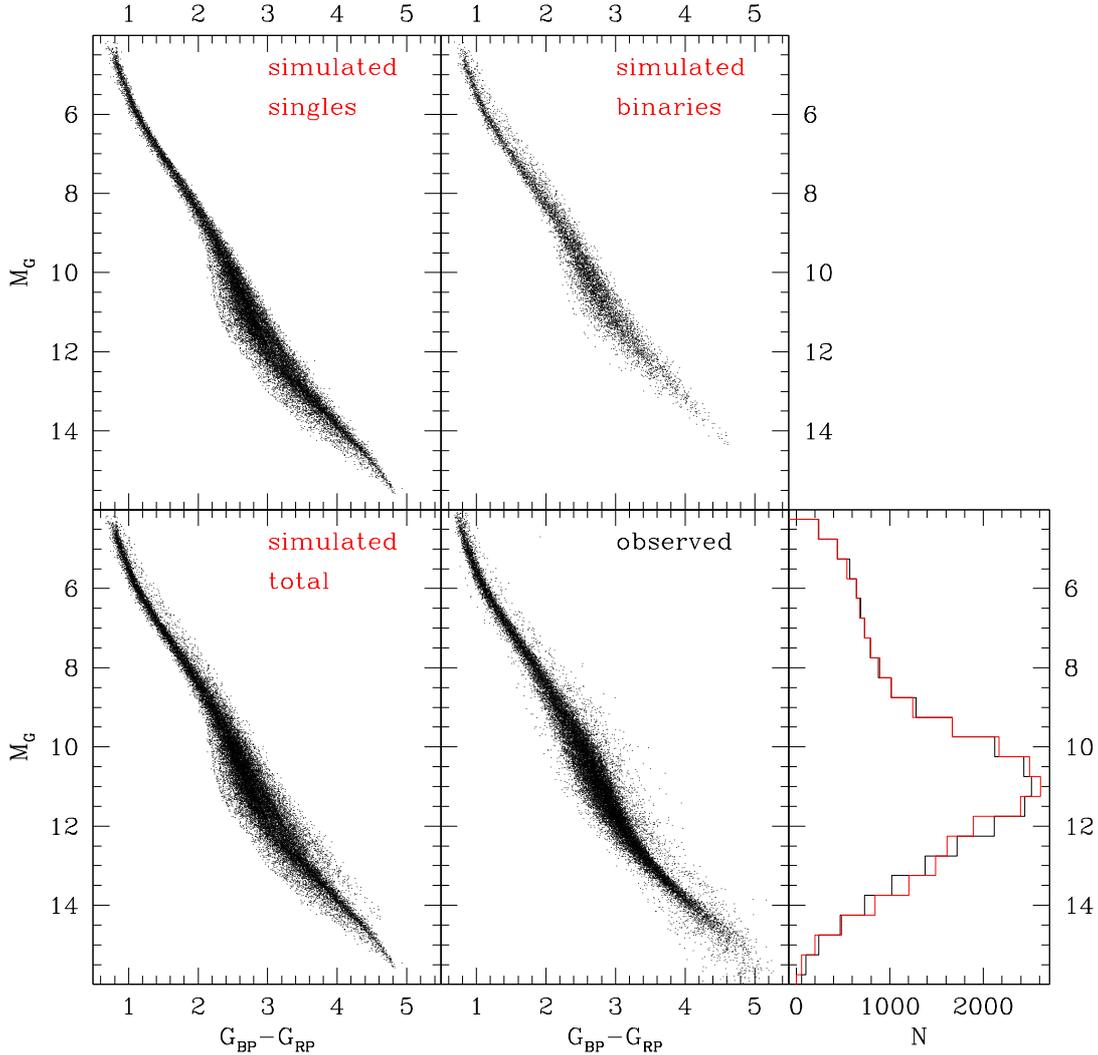}
 \caption{Comparison between the synthetic (bottom-left panel) and observed
 (bottom-central panel) CMD of the {\it nearby 
 sample}. The contributions of single stars (top-left panel) and binaries
 (top-central panel) to the synthetic CMD 
 are shown separately. The observed (black) and synthetic (red) $G$-band
 luminosity functions are compared in the bottom-right panel.}
\label{sim}
\end{figure*}

The algorithm adopted in this analysis is based on the best fit of the
distribution of stars in the $(G_{BP}-G_{RP})-M_{G}$ colour-absolute magnitude 
diagram (CMD) with a synthetic stellar population.

The masses of $10^{6}$ synthetic particles were extracted from a MF defined
in 20 evenly-spaced mass intervals of 0.045 $M_{\odot}$ width, from 0.1 to 1 $M_{\odot}$.
The relative fraction of stars in each mass bin ($m_{i}$) is set by 20 coefficients ($k_{i}$)

For a given MF, synthetic absolute magnitudes and colours were derived by 
interpolating through the set of isochrones of the MESA database \citep{2016ApJ...823..102C} adopting an
age of 10 Gyr. This choice is justified by the fact that the {\it nearby sample} 
is formed by MS stars below the turn-off. In the assumption of a constant 
star formation rate, 98\% of them are older than 200 Myr i.e. the timescale
needed by a 0.2 $M_{\odot}$ star to
end its pre-MS phase \citep{2011A&A...533A.109T}. Under these conditions, evolutionary effects are 
negligible and single-age old isochrones properly reproduce the mass-luminosity
relation of these stars.

The metallicity distribution was modelled as an asymmetric gaussian with 
mode [Fe/H]=0 and a standard deviation of the metal-poor tail 
$\sigma_{Fe,low}=0.2~dex$, while the standard deviation of the metal-rich tail
$\sigma_{Fe,hi}$ was left as a free parameter.
This model provides the best fit to the colour distribution of the {\it
nearby sample} and agrees with the metallicity distribution derived by the
spectroscopic analysis of 4666 stars in the solar neighbourhood by \citet{2017A&A...600A..22M}.

Synthetic particles were distributed at different heights above the 
Galactic plane following a gaussian distribution with $\sigma_{Z}=147~pc$ and
homogeneously along the direction parallel to the Galactic plane over a volume
twice larger than that defined for the {\it nearby sample}.
The value of $\sigma_{Z}$ was chosen as the one maximizing the log-likelihood

\begin{equation*}
ln L=-N~ln~C+\sum_{i=1}^{N} ln~P_{i}
\end{equation*}

where

\begin{equation*}
P_{i}=\int_{-\infty}^{\infty} exp\left[-\frac{(p-p_{i})^{2}}{2\epsilon_{i}^{2}}-\frac{(p^{-1} sin~b_{i}+Z_{\odot})^{2}}{2
\sigma_{Z}^{2}}\right] dp
\end{equation*}

\begin{multline*}
C=\sqrt{\frac{\pi}{2}}~\sigma_{Z}(r_{max}^{2}-Z_{\odot}^{2}-\sigma_{Z}^{2})\times\\
\left[Erf\left(\frac{r_{max}+Z_{\odot}}{\sqrt{2}\sigma_{Z}}\right)+
Erf\left(\frac{r_{max-Z_{\odot}}}{\sqrt{2}\sigma_{Z}}\right)\right]+\\
\sigma_{Z}^{2}\left\{(r_{max}-Z_{\odot})~exp\left[-
\frac{(r_{max}+Z_{\odot})^{2}}{2\sigma_{Z}^{2}}\right]+\right.\\
\left.(r_{max}+Z_{\odot})~
exp\left[-\frac{(r_{max}-Z_{\odot})^{2}}{2\sigma_{Z}^{2}}\right]\right\}
\end{multline*}

in these equations N is the number of stars in the {\it nearby sample}, 
$b_{i},~p_{i}$ and $\epsilon_{i}$ are the Galactic latitude, the parallax
and its associated error of the i-th star, $r_{max}=50~pc$ and
$Z_{\odot}=1.4~pc$ (see Sect.
\ref{data_sec}).

The population of binaries was simulated by random pairing a fraction 
$f_{b}$ of synthetic stars whose fluxes in the $G$, $G_{BP}$ and $G_{RP}$
passbands were summed.

The dereddened colours and absolute magnitudes of synthetic particles were converted
into apparent ones by inverting eq. \ref{abs_eq} and a real star with $G$ magnitude within
0.25 mag was associated to each synthetic particle. A gaussian shift in 
colour, magnitude and parallax with standard deviation equal to
the uncertainty of the associated star in the corresponding quantity was
then added defining an "observational" set of synthetic apparent magnitudes and
parallaxes. Stars with "observational" magnitudes and parallaxes outside
the adopted cuts (see Sect. \ref{data_sec}) were rejected. 
The "observational" absolute magnitude of synthetic stars was then re-computed
using eq. \ref{abs_eq}. The above task mimics the effect of photometric and 
astrometric errors and ensures a proper treatment of the asymmetric distance 
error \citep{2018A&A...616A...9L}.

For an assumed binary fraction, an iterative algorithm was used to
determine the best fit MF.
At the first iteration, guess values of the coefficients $k_{i}\propto m_{i}^{-2.35}$ 
and of $\sigma_{Fe,hi}=0.2~dex$ were adopted. 
The $(G_{BP}-G_{RP})-M_{G}$ CMD was divided
in 20 bins defined to include stars lying in a colour range within 3$\sigma$ 
about the MS ridge line and at magnitudes corresponding to the mass range of the same 
20 bins defining the MF (see above) on a 10 Gyr-old isochrone with solar metallicity. 
An additional sample of stars with colours redder than
3$\sigma$ with respect to the MS mean ridge line was defined to include
unresolved binary systems (see Fig \ref{cmd}).
The number of {\it nearby sample} stars ($N_{obs,i}$) and synthetic particles ($N_{synth,i}$) contained in
each bin were counted and the coefficients $k_{i}$ were updated using the following correction

\begin{equation*}
k_{i}'=k_{i} \frac{N_{obs,i}\sum_{i} N_{synth;i}}{N_{synth,i}\sum_{i} N_{obs,i}}
\end{equation*}

The value of $\sigma_{Fe,hi}$ was also updated, by increasing (decreasing) its 
value by 0.005 dex if the fraction of {\it nearby sample} stars contained in the 
binary region was larger (smaller) than that in the synthetic CMD.
The above procedure was repeated until convergence, 
providing for each choice of $f_{b}$ a set of best fit coefficients $k_{i}$.
The difference between the distribution of observed and synthetic stars in 
the CMD was quantified using the penalty function

\begin{equation}
\xi=-\sum_{j} ln~\rho_{j}
\label{chi_eq}
\end{equation}
where $\rho_{j}$ is the density of synthetic particles in the CMD at the position of the j-th 
{\it nearby sample} star
\begin{equation*}
\rho_{j}=\left[\left(\frac{(G_{BP}-G_{RP})_{10}-(G_{BP}-G_{RP})_{j}}{\Delta_{BR}}\right)^{2}+\left(\frac{G_{10}-G_{j}}{\Delta_{G}}\right)^{2}\right]^{-1}
\end{equation*}

where $(G_{BP}-G_{RP})_{j},~g_{j},~(G_{BP}-G_{RP})_{10}$ and $G_{10}$ are
the colours and magnitudes of the j-th {\it nearby sample} star and of its 10-th
nearest neighbour synthetic particle, respectively, and $\Delta_{BR}$ and $\Delta_{G}$ define
the metric in the CMD. The best metric is the one maximizing
the entropy in the CMD so that
$\Delta_{G}/\Delta_{BR}=\sigma_{G}/\sigma_{BR}\sim 2$
where $\sigma_{G}$ and $\sigma_{BR}$ are the standard deviations of magnitude
and colour in the {\it nearby sample}.
The comparison between the CMD and $G$-band luminosity function of the {\it
nearby sample} and those of the best fit synthetic model is shown in Fig.
\ref{sim}.

Uncertainties were estimated through a Monte Carlo technique: at each step a
synthetic CMD containing the same number of stars of the {\it nearby sample} 
was simulated assuming the best fit MF, metallicity distribution and binary fraction and
its MF was estimated in the same fashion as for real data. The r.m.s of
the MFs of $10^{4}$ different simulations were adopted as the corresponding
uncertainties. This procedure takes into account the effect of Poisson noise
but does not include the effect of all the systematics (e.g. uncertainties in
isochrones, mass-ratio distribution of binaries, spatial distribution, limiting
magnitude, etc.).

\subsection{Results}
\label{reslo_sec}

\begin{table}
 \centering
  \caption{PDMF of the solar neighbourhood in the sub-solar mass regime.
  The determinations using two different stellar evolution models are listed. 
  In both cases, the adopted magnitude interval is $7.5<G<$18.}
  \begin{tabular}{@{}lcccc@{}}
  \hline
                   &\multicolumn{2}{c}{MESA}   & \multicolumn{2}{c}{PARSEC}\\
   $f_{b}$         &\multicolumn{2}{c}{25\%}   & \multicolumn{2}{c}{30\%}\\
   $\sigma_{Fe,hi}$ &\multicolumn{2}{c}{0.13}  & \multicolumn{2}{c}{0.14}\\
 log $M/M_{\odot}$ & log $\psi$ & $\epsilon_{log \psi}$ & $log \psi$ & $\epsilon_{log
 \psi}$\\
 \hline
-0.953 & 1.03 & 0.03 & 1.10 & 0.01\\
-0.811 & 1.22 & 0.01 & 1.08 & 0.02\\
-0.704 & 1.08 & 0.03 & 1.00 & 0.02\\
-0.619 & 0.86 & 0.05 & 0.91 & 0.03\\
-0.547 & 0.79 & 0.04 & 0.82 & 0.05\\
-0.486 & 0.62 & 0.04 & 0.71 & 0.03\\
-0.432 & 0.50 & 0.04 & 0.64 & 0.06\\
-0.385 & 0.42 & 0.05 & 0.57 & 0.06\\
-0.342 & 0.36 & 0.03 & 0.51 & 0.06\\
-0.302 & 0.26 & 0.06 & 0.43 & 0.06\\
-0.266 & 0.32 & 0.03 & 0.39 & 0.07\\
-0.233 & 0.26 & 0.04 & 0.33 & 0.06\\
-0.202 & 0.23 & 0.05 & 0.29 & 0.07\\
-0.174 & 0.25 & 0.04 & 0.31 & 0.05\\
-0.147 & 0.15 & 0.05 & 0.22 & 0.05\\
-0.121 & 0.08 & 0.05 & 0.15 & 0.06\\
-0.097 & 0.13 & 0.05 & 0.12 & 0.07\\
-0.074 & 0.08 & 0.08 & 0.02 & 0.05\\
-0.053 & 0.18 & 0.05 & 0.16 & 0.10\\
-0.032 & 0.08 & 0.18 & 0.08 & 0.11\\
\hline
\end{tabular}
 \label{tab:table1}
\end{table}

\begin{figure*}
 \includegraphics[width=15cm]{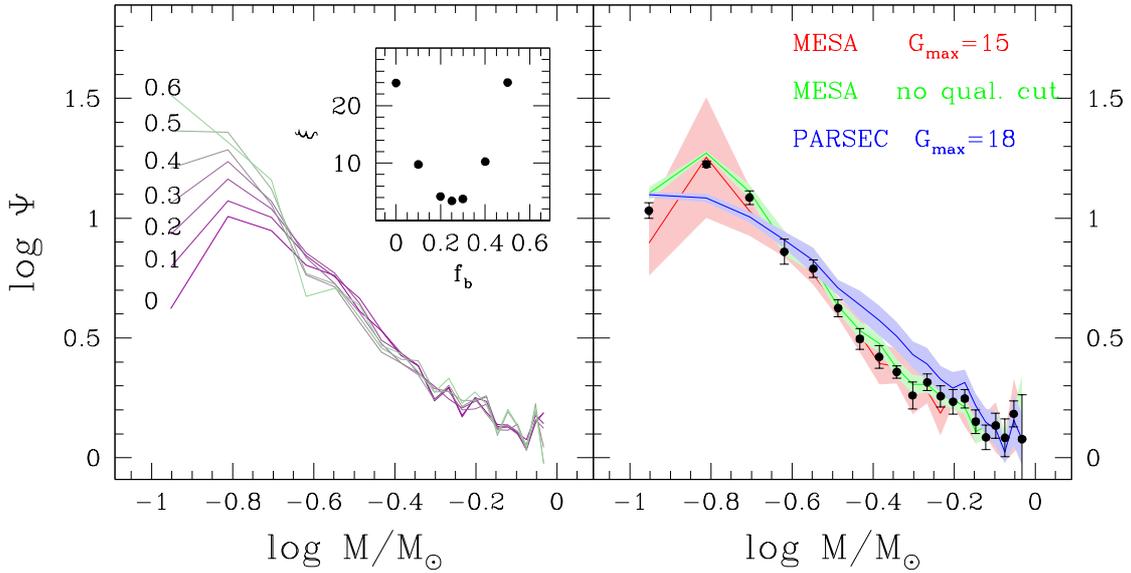}
 \caption{Left panel: PDMF of the solar neighbourhood in the low-mass regime
 ($M<1~M_{\odot}$) for different assumptions of the binary fraction $f_{b}$. The inset
 shows the behaviour of the penalty function $\xi$ as a function of $f_{b}$. Right panel:
 Comparison between the MF derived using different sets of isochrones, quality cuts and
 limiting $G$ magnitudes. The shaded area indicate the
 1$\sigma$ uncertainties. All the MFs are normalized to their values at
 $1~M_{\odot}$.}
\label{fbmf}
\end{figure*}

The best fit MFs for different adopted binary fractions and their corresponding 
values of $\xi$ are shown in the left panel of Fig. \ref{fbmf}. 
It is apparent that the slope of the MF in the mass range $0.25<M/M_{\odot}<1$ is 
almost independent on the adopted binary fraction, being nicely fit by a 
single power-law with an index ranging from $\alpha=-1.38$ 
to -1.16 for $10\%<f_{b}<60\%$, and a best fit value of $\alpha=-1.34\pm0.07$ at 
$f_{b}=25\%$. 
At lower masses the MF significantly flattens and has a peak at 
$M\sim 0.15~M_{\odot}$ although its shape strongly depends on the adopted binary fraction.

The best fit binary fraction ($f_{b}=25\%$) is lower 
than that measured by the long-baseline spectroscopic campaigns performed 
in the past \citep[$\sim50\%$;][]{1991A&A...248..485D,2017ApJS..230...15M}. Note however that the fraction of binaries estimated 
here is quite uncertain ($\epsilon_{f_{b}}\sim10\%$) and refers to unresolved binaries, while some of 
the wide binaries at small heliocentric distances are resolved by Gaia and therefore 
included in the {\it nearby sample} as single stars. Moreover, the selection on astrometric 
quality described in Sect. \ref{data_sec} can potentially exclude those binaries for which the relative 
motion of their components alters the position measured by Gaia, thus worsening the quality of the fit (see below).
The corresponding metallicity dispersion on the metal-rich side turns out to be
$\sigma_{Fe,hi}=0.13~dex$, in agreement with the result by \citet{2017A&A...600A..22M}. 

To test the dependence of the measured MF from other assumptions made in the analysis, 
I repeated the above procedure {\it i)} using the isochrones from the PARSEC database
\citep{2012MNRAS.427..127B}, 
{\it ii)} assuming a limiting magnitude of $G<15$, and {\it iii)} removing the 
quality cut in the Gaia astrometric solution (see the right panel of 
Fig.\ref{fbmf}). The mean slope of the MF does not depend on the adopted isochrones 
($\alpha_{PARSEC}=-1.51\pm 0.07$), while some small-scale differences are apparent 
due to differences in the mass-luminosity relation of these two models.
For example, the steepening at $M<0.5~M_{\odot}$ apparent when
using MESA isochrones is absent in the MF calculated using PARSEC models which are
likely spurious. Moreover, with this latter set of models, the MF at 
$M<0.15~M_{\odot}$ is almost flat and does not show any clear peak.
It is also possible to fit the MF calculated using the PARSEC isochrones in 
this mass range with a log-normal
function with central value $log~(M_{0}/M_{\odot})=1.06$ and $\sigma_{log
M}=0.44$, while this analytical representation provides a poor fit at masses
$M<0.2~M_{\odot}$ when MESA isochrones are used.
The MFs in this regime derived using the two sets of models mentioned above are listed in
Table \ref{tab:table1}.
No significant differences are noticeable by either changing the adopted limiting 
magnitude or removing the selection on astrometric quality, indicating 
that the completeness at $15<G<18$ is still high and that the fraction of artifacts is small 
and homogeneously distributed in magnitude. It is however worth noting that, 
when no selection cut on astrometric quality is applied, the best fit is obtained with a 
fraction of binaries $f_{b}=40\%$, significantly higher than that obtained in the selected sample, indicating that a 
sizeable fraction of binaries is rejected by the quality cut. 
This explains the discrepancy between the fraction of binaries estimated here and that of previous 
literature works.

Since all the stars of the solar neighbourhood in the sub-solar mass regime did not have
enough time to evolve off the MS, and because of the non-collisional nature of
the Galactic disk, the above derived PDMF is representative of the IMF.

\section{High-mass regime}
\label{hi_sec}

\subsection{Method}
\label{methi_sec}

\begin{figure*}
 \includegraphics[width=15cm]{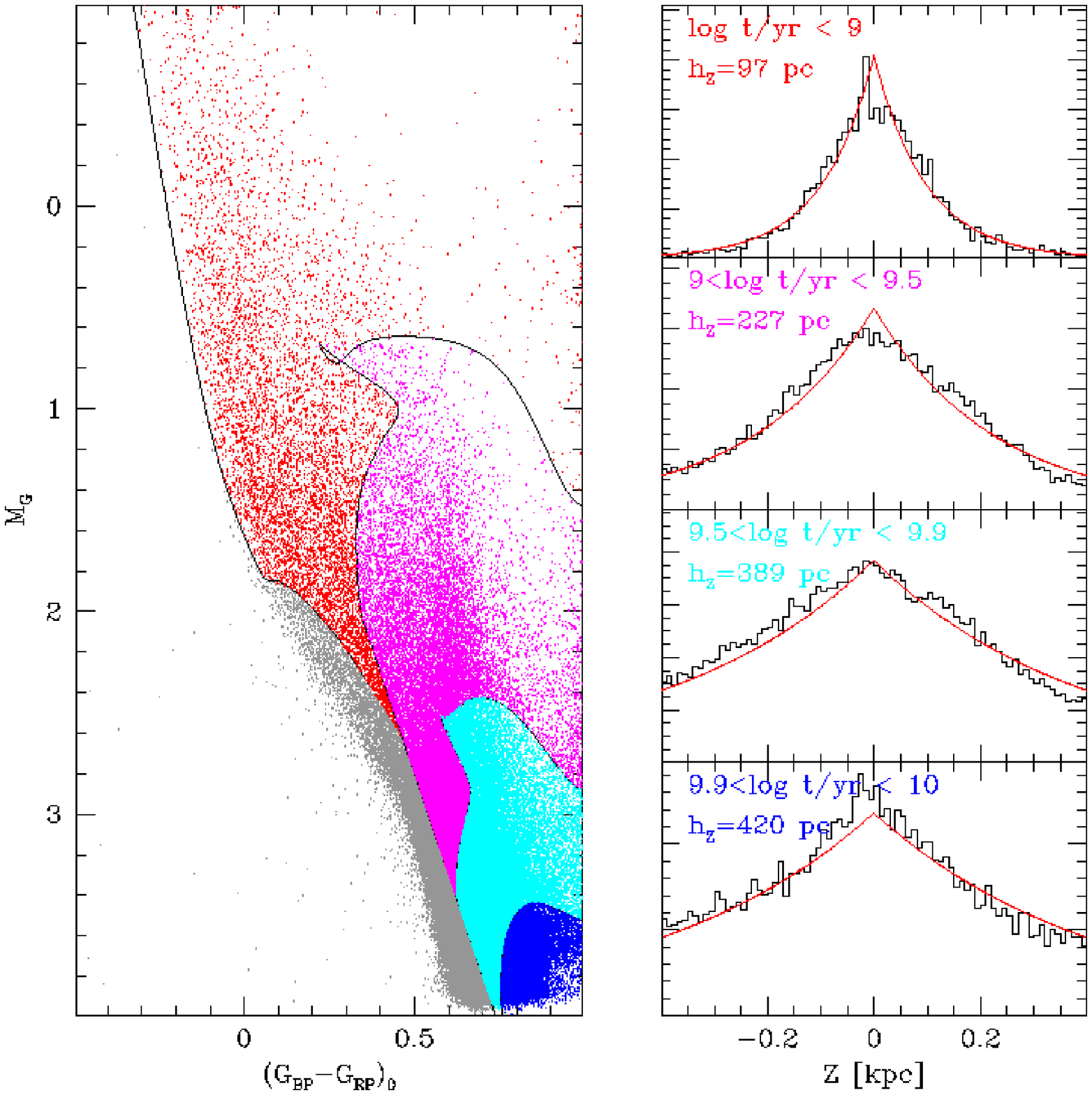}
 \caption{Left panel: CMD of the {\it bright sample} without the upper magnitude
 cut at $G=7.5$. The selection boxes of the four age bins are marked by black
 lines. Selected stars are plotted with red ($log~t/yr<9$), magenta
 ($9<log~t/yr<9.5$), cyan ($9.5<log~t/yr<9.9$) and blue ($9.9<log~t/yr<10$) dots.    
 Right panels: histograms of the height above the Galactic plane of the stars
 belonging to the four age bins. The best fit exponential function is marked by
 the red line in each panel.}
\label{vartz}
\end{figure*}

\begin{figure*}
 \includegraphics[width=15cm]{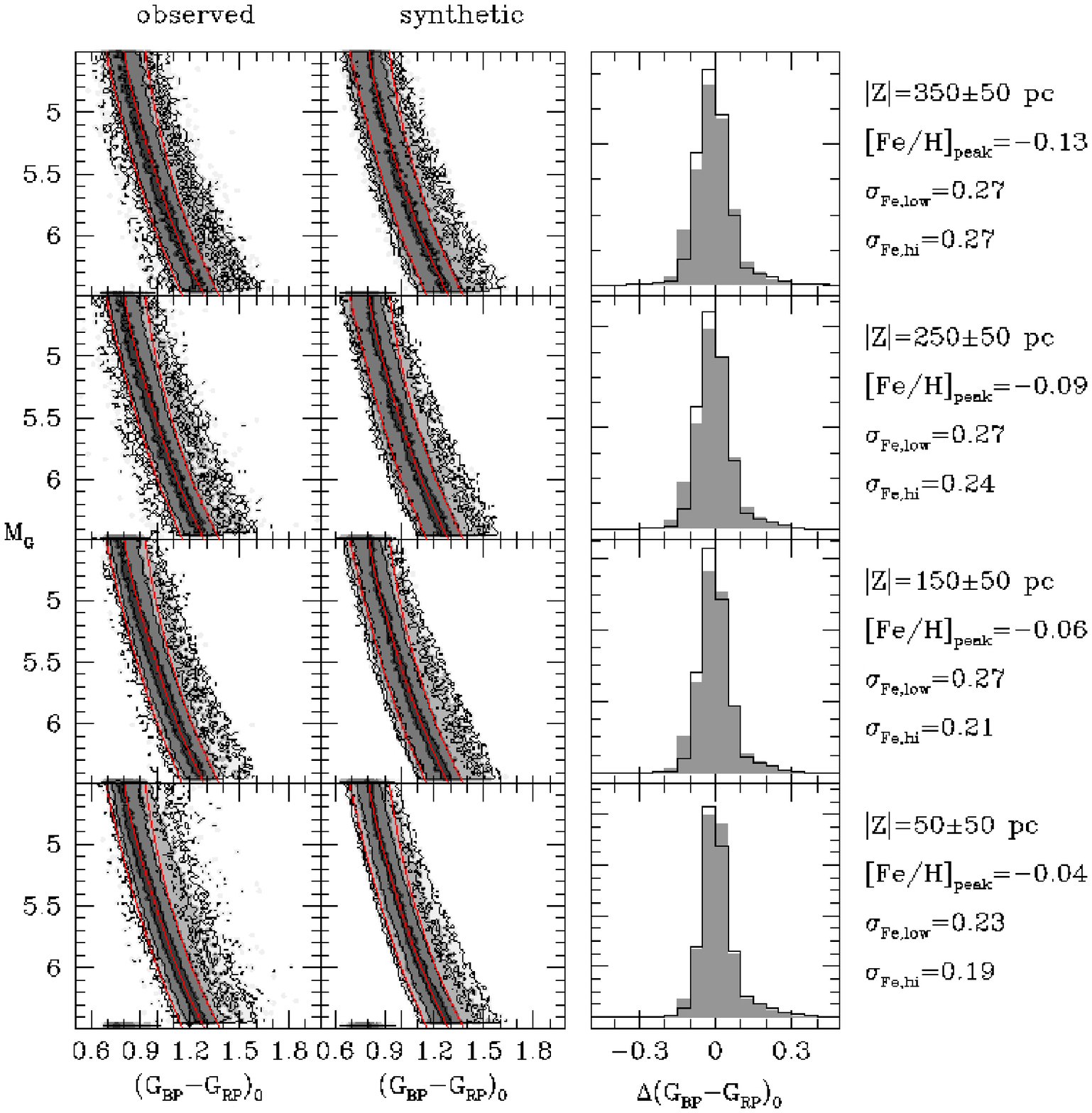}
 \caption{Observed (left panels) and synthetic (middle panels) densities of stars in the CMD of the solar 
 neighbourhood in the absolute magnitude range $4.5<M_{G}<6.5$ in slices at 
 different heights above the Galactic plane. Darker contours delimit regions with higher densities 
 increasing in logarithmic steps of 0.5 dex. The MESA 10 Gyr-old isochrones 
 with metallicities [Fe/H]=-0.5, 0 and +0.5 (from left to right) are overplotted as red lines. 
 The distributions of colour differences about the isochrone at
 solar metallicity are shown in the right panels for the observed and synthetic 
 CMDs with empty and grey histograms, respectively.}
\label{varfez}
\end{figure*}

At odds with the portion of CMD fainter than the turn-off point (i.e. less massive than the
oldest star which exhausted hydrogen at its centre), the bright part of
the CMD is mainly populated by those stars with ages smaller than their
evolutionary timescales (see Sect. \ref{intro_sec}). In this situation the PDMF
differs from the IMF which can be estimated only assuming a SFH.
This last function can be determined on the basis of the overall distribution of
stars in the CMD \citep{1989Ap&SS.156....9R}.

For this purpose I simulated the CMD of the {\it bright sample} as the
superposition of stellar populations with different ages. The number and width of 
the age bins determining the resolution of the derived SFH should be chosen as a 
compromise to ensure flexibility while limiting the degeneracy caused by the increasing number 
of free parameters.
Two cases were considered: {\it i)} a low-resolution SFH defined by 4 stellar 
populations (with ages $log~t/yr<9,~9<log
~t/yr<9.5,~9.5<log~t/yr<9.9$ and $9.9<log~t/yr<10$) and {\it ii)} a high-resolution SFH 
defined by 10 stellar populations with ages evenly spaced from 0 to 10 Gyr with a width of 1 Gyr.
Within each age bin, star ages were randomly extracted so that they evenly
populate the bin. The age upper limit was chosen from the comparison
between the lower envelope of the Subgiant Branch observed in the CMD of the 
{\it bright sample} with the MESA isochrone with suitable metallicity (see below).

It is well known that stars at different heights above the Galactic plane have 
different ages and metallicity distributions. This is a consequence of the 
dependence of the star formation efficiency on the density, on the secular 
evolution of the vertical distribution of stars and of the increasing
contamination of the thick disc (see Sect. \ref{intro_sec}). 
To account for these effects the vertical variations of the age and metallicity
distributions were modelled.
In particular, the density of stars in each age bin was fitted with an
exponential function with scale-height $h_{Z}(t)$. To determine the appropriate 
scale-height, a synthetic CMD assuming a constant SFR and a 
\citet{2001MNRAS.322..231K} IMF\footnote{The synthetic CMD simulated 
in this task is used only to define the selection box used to compute the vertical 
scale-height of stellar populations as a function of their ages. The adopted 
shape of the IMF as well as the adopted binary fraction and SFR have almost no 
impact on the final result.} was simulated and a selection 
box in the CMD where the fraction of synthetic
particles in the considered age interval is $>90\%$ was defined. 
The {\it bright sample} stars comprised within the appropriate selection box of
the dereddened colour-absolute magnitude diagram were
used to search the value of $h_{Z}$ which maximizes the log-likelihood

\begin{multline*}
ln L=-N~ln~h_{Z}-N~ln\left[ 1-exp\left(-\frac{Z_{max}}{h_{Z}}\right)\right]+\\
\sum_{i=1}^{N} ln~\int_{-\infty}^{\infty}
exp\left[-\frac{(p-p_{i})^{2}}{2 \epsilon_{i}^{2}}-\frac{|p^{-1}
sin~b_{i}+Z_{\odot}|}{h_{Z}}\right] dp
\end{multline*}

where $Z_{max}=390~pc$ and $Z_{\odot}=1.4~pc$ (see Sect. \ref{data_sec}).
Since the bright magnitude cut at $G=7.5$ removes most of the bright stars at 
small heliocentric distances altering the overall shape of the
distribution, I relaxed this criterion only for this task, thus including
all stars brighter than $G<18$. Note that the (possible) incompleteness at bright magnitudes
affects only the brightest stars in the youngest age bin at small distances,
while the fit is driven by the tails of the distribution. Therefore, this
exception is not expected to affect the final result. 
As expected, the best fit scale-heights for the corresponding age bins 
increase with age (see Fig. \ref{vartz}).
For each age bin, synthetic particles were distributed at different heights
above the Galactic plane according to the corresponding distribution and  
homogeneously along the direction parallel to the Galactic plane over a volume
twice larger than that defined for the {\it bright sample}.

The metallicity distribution at different heights above the Galactic plane was
estimated by best-fitting the colour distribution of MS stars
($4.5<M_{G}<6.5$) selected 
from the Gaia catalog in 4 slices at different heights $\langle
|Z|\rangle$ from 50 to 350 pc with a 100 pc width.
The absolute magnitudes and dereddened $G_{BP}-G_{RP}$ colours of these stars
were calculated using eq.s \ref{abs_eq} and \ref{redd_eq}.
A synthetic CMD of that portion of the CMD was simulated using the technique
described in Sect. \ref{metlo_sec} and the best fit binary fraction 
$f_{b}=25\%$ derived in the {\it nearby} sample (appropriated for these 
low-mass stars).
In each slice, the metallicity distribution was modelled as an asymmetric
gaussian characterized by a mode ($[Fe/H]_{peak}$) and two different standard deviations
at the two sides of the distribution ($\sigma_{Fe,low}$ and $\sigma_{Fe,hi}$). 
The values of these parameters minimizing the penalty function
of eq. \ref{chi_eq} were chosen as representative of the considered slice.
At increasing heights above the Galactic plane, the 
metallicity distribution appears to shift toward the
metal-poor range becoming more symmetric and with increasing dispersions at both
sides (see Fig. \ref{varfez}).
It is worth noting that the derived metallicity variations are relatively small. 
This is not surprising since the expected contamination from thick disc stars 
is small: a comparison with the \citet{2003A&A...409..523R} model suggests that 
only 3.3\% of the {\it bright} sample stars should belong to the thick disc.
The metallicity of each star was then extracted by linearly interpolating
through the defined distributions according to its height above the
Galactic plane. The assumptions made above naturally introduce a height-dependent
age-metallicity relation with the particles at larger heights being on average
older and more metal-poor than those close to the Galactic plane.

The magnitudes and colours of synthetic stars were derived by interpolating
through the set of MESA isochrones of appropriate age and metallicity and
assuming a broken power-law IMF with index $\alpha=-1.34$ at $M<1~M_{\odot}$
(see Sect. \ref{reslo_sec}) and leaving the slope in the high-mass slope as a free
parameter. Particles with ages higher than the evolutionary timescales associated
to their masses and metallicities
were automatically removed from the sample.

A population of binaries was also simulated by adding to the $G,~G_{BP}$
and $G_{RP}$ fluxes of a fraction $f_{b}$ of stars those of companion stars with
mass extracted from the distribution of mass-ratios described by
\citet{2017ApJS..230...15M} for A-type stars. To limit the number of free
parameters, a fixed value of $f_{b}=50\%$ was adopted \citep{1991A&A...248..485D}.
This choice was made to account for the fast changing fraction
of multiple systems found by \citet{2017ApJS..230...15M} in this mass range (50\%-90\%) and 
considering that a small fraction of binaries could be actually resolved in the {\it bright} sample.
The effect of different fractions of binaries is tested in Sect. \ref{reshi_sec}.

The effect of reddening, photometric and parallax errors was simulated using the same technique
described in Sect. \ref{metlo_sec} and particles satisfying the magnitude
and positional cuts adopted for the {\it bright sample} (see Sect.
\ref{data_sec}) are retained.

The relative fractions of stars in the age bins were estimated by matching the
$G$-band luminosity function of MS stars at $G_{BP}-G_{RP}<1$ using a
least-squares fitting algorithm providing the SFH. Because of the degeneracy between MF
slope and SFH, it is always possible to reproduce the luminosity function of MS 
stars for any choice of the high-mass MF slope. On the other hand, the
proportion of old stellar populations is constrained by the relative
fraction of evolved stars populating the red portion of the CMD along the Red
Giant Branch. The penalty function defined in eq. \ref{chi_eq} was
calculated using all the stars in the CMD of the {\it bright sample} and the
value of the high-mass IMF slope providing the lowest value of $\xi$ was
derived.

The uncertainty attached to the MF slope was calculated using a Monte
Carlo technique (see Sect. \ref{metlo_sec}). Although the estimated uncertainty is relatively
small, the error budget is dominated by systematics. Indeed, many priors
were assumed in the above analysis each of them significantly affecting the
final estimate.

Because of the relatively bright cut adopted for our sample ($G>7.5$) massive
stars are poorly represented: while the maximum mass in the best
fit synthetic catalog is $\sim 10~M_{\odot}$, the strongest constraint to the IMF
slope is given by stars with $1<M/M_{\odot}<2.5$ which represent more than 96\% of
the sample. This last interval was considered as the range of validity of
the present analysis. 

\subsection{Results}
\label{reshi_sec}

\begin{table}
 \centering
  \caption{IMF of the solar neighbourhood in the super-solar mass regime.
  The determinations using two different SFH resolutions are listed. 
  All the determinations are made using MESA isochrones, except the last line where 
  PARSEC isochrones are used.}
  \begin{tabular}{@{}lcccccc@{}}
  \hline
                 &			    &		  & \multicolumn{2}{c}{low-res SFH} & \multicolumn{2}{c}{hi-res SFH}\\
                 &	                    &	qual.	  & \multicolumn{2}{c}{4 bins}      & \multicolumn{2}{c}{10 bins}\\
   $f_{b}$  (\%) &  $\alpha_{<1~M_{\odot}}$ &   flag      & $\alpha$  & $\epsilon_{\alpha}$ & $\alpha$	& $\epsilon_{\alpha}$\\
 \hline
30 	      	 & -2			   & yes	  & -2.55 & 0.08 & -2.32 & 0.10\\
50 	      	 & -2			   & yes	  & -2.70 & 0.09 & -2.41 & 0.13\\
70 	      	 & -2			   & yes	  & -2.74 & 0.08 & -2.46 & 0.10\\
50 	      	 & -2			   & no 	  & -2.58 & 0.09 & -2.23 & 0.12\\
30 	      	 & -1.34		   & yes	  & -2.57 & 0.06 & -2.33 & 0.10\\
50 	      	 & -1.34		   & yes	  & -2.68 & 0.09 & -2.41 & 0.11\\
70 	      	 & -1.34		   & yes	  & -2.74 & 0.07 & -2.46 & 0.09\\
50 	      	 & -1.34		   & no 	  & -2.57 & 0.07 & -2.24 & 0.12\\
30 	      	 & -0.5 		   & yes	  & -2.59 & 0.07 & -2.32 & 0.11\\
50 	      	 & -0.5 		   & yes	  & -2.67 & 0.11 & -2.42 & 0.11\\
70 	      	 & -0.5 		   & yes	  & -2.74 & 0.08 & -2.45 & 0.11\\
50 	      	 & -0.5 		   & no 	  & -2.57 & 0.07 & -2.25 & 0.13\\
50 (PARSEC)      & -1.34		   & yes	  & -4.05 & 0.15 & -3.41 & 0.14\\
\hline
\end{tabular}
 \label{tab:table2}
\end{table}

\begin{figure*}
 \includegraphics[width=15cm]{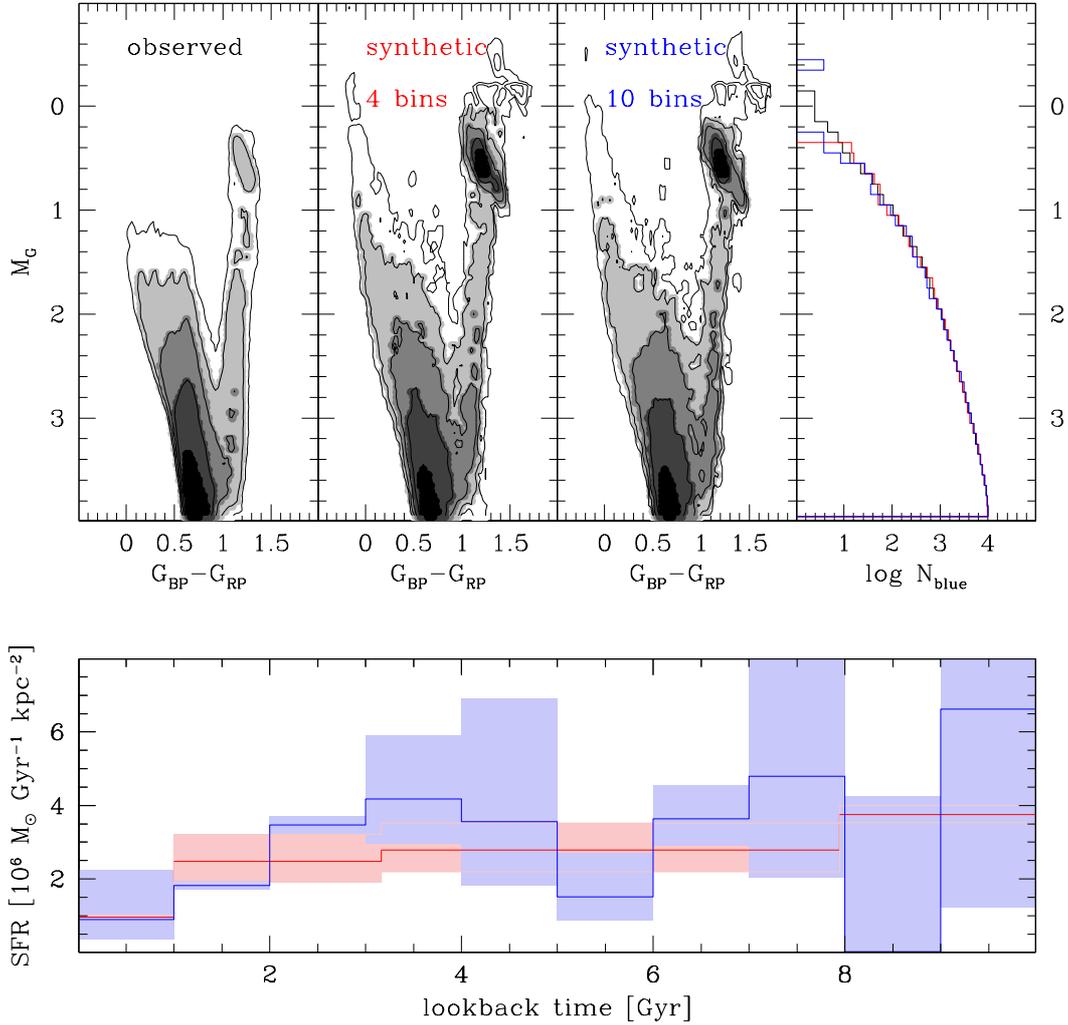}
 \caption{Top panels: comparison between the density in the CMD of the {\it bright sample}
 (top-left) and
 the prediction of the best fit composite stellar population using 4 (top middle-left) and 10 age components (top middle-right), respectively. 
 Darker contours in the left and central panels delimit regions with higher
 particle densities increasing by logarithmic steps of 0.5 dex. In the top-right
 panel the observed and synthetic $G$-band luminosity function of stars bluer
 than $G_{BR}-G_{RP}<1$ are shown with black, red and blue lines, respectively. Bottom
 panel: output SFR of the best fit models. The shaded areas indicate the 1$\sigma$ 
 uncertainty.}
\label{bright}
\end{figure*}

As a result of the procedure described in the previous section, adopting a low-resolution SFH 
the best fit IMF slope in the super-solar mass regime turns out to be $\alpha=-2.68\pm 0.09$.
The best fit model to the {\it bright sample} CMD and the $G$-band luminosity
function are shown in Fig. \ref{bright} together with the derived SFH. 
The SFH is almost constant over the past 10 Gyr with a slight decrease at recent
epochs, in agreement with the results by 
\citet{2018IAUS..330..148B} and \citet{2018A&A...620A..79M} within the uncertainties. 

A significantly different result is obtained when the high-resolution SFH is adopted.
The resulting best fit CMD, $G$-band luminosity
function and SFH are also shown in Fig. \ref{bright}.
The corresponding MF turns out to be $\alpha=-2.41\pm 0.11$, significantly flatter 
than that derived using fewer age components.
It can be noticed that, while the general trend of the SFH is compatible with 
that obtained in the low-resolution case with an increasing star formation rate at old ages, 
the SFH appears more bursty, being characterized by three intense 
episodes of star formation which occurred 3, 7 and 10 
Gyr ago. However, given the large number of free parameters, it is not clear if 
such a behaviour is real or rather it is due to the noise or to the degeneracy between 
the various components.

As reported in Sect. \ref{methi_sec}, beside the effect of the SFH, the results for this mass range strongly
depend on the various assumptions.
The strongest effect is produced by the adopted isochrones: by 
repeating the analysis using PARSEC isochrones, the derived IMF slope steepens 
to $\alpha<-4$. This is a consequence of the different assumptions made by these
models on the overshooting in low-mass stars affecting the time spent along the
Red Giant Branch and therefore star counts in this evolutive sequence.

The dependence of the derived MF slope as a function of the 
binary fraction was checked by repeating the analysis assuming different 
values of $f_{b}$.
The high-mass MF slope is found to mildly depends on this assumption varying 
by $\Delta\alpha=\pm 0.08$ for binary fractions from 30\% 
to 70\%. This variation is of the order of the random error 
so that uncertainties in the prescriptions for the population of binaries are 
not expected to significantly affect the MF slope in this mass range.

The high-mass MF slope is also found to be independent on the adopted slope at 
low-masses: by assuming a value in the range $-2<\alpha~(M<1~M_{\odot})<-0.5$ the 
derived slope in the high-mass range changes by $\Delta \alpha<0.01$. This is not surprising 
since the {\it bright} sample has been specifically designed to contain only 
stars at $M_{G}<4$ where only a few low-mass stars in the metal-poor tail of the 
metallicity distribution are present.

The MF slope was also calculated 
without applying any selection on the astrometric quality parameter (see Sect. \ref{data_sec}). 
In this case the slope of the MF flattens slightly, remaining however 
compatible within the errors with that derived for the selected sample.

The entire set of high-mass IMF slopes derived under various assumptions and their 
associated uncertainties are summarized in Table 2. 

\section{comparison with previous IMF determinations}
\label{comp_sec}

\subsection{Solar neighbourhood}
\label{comp_solar_sec}

\begin{figure*}
 \includegraphics[width=15cm]{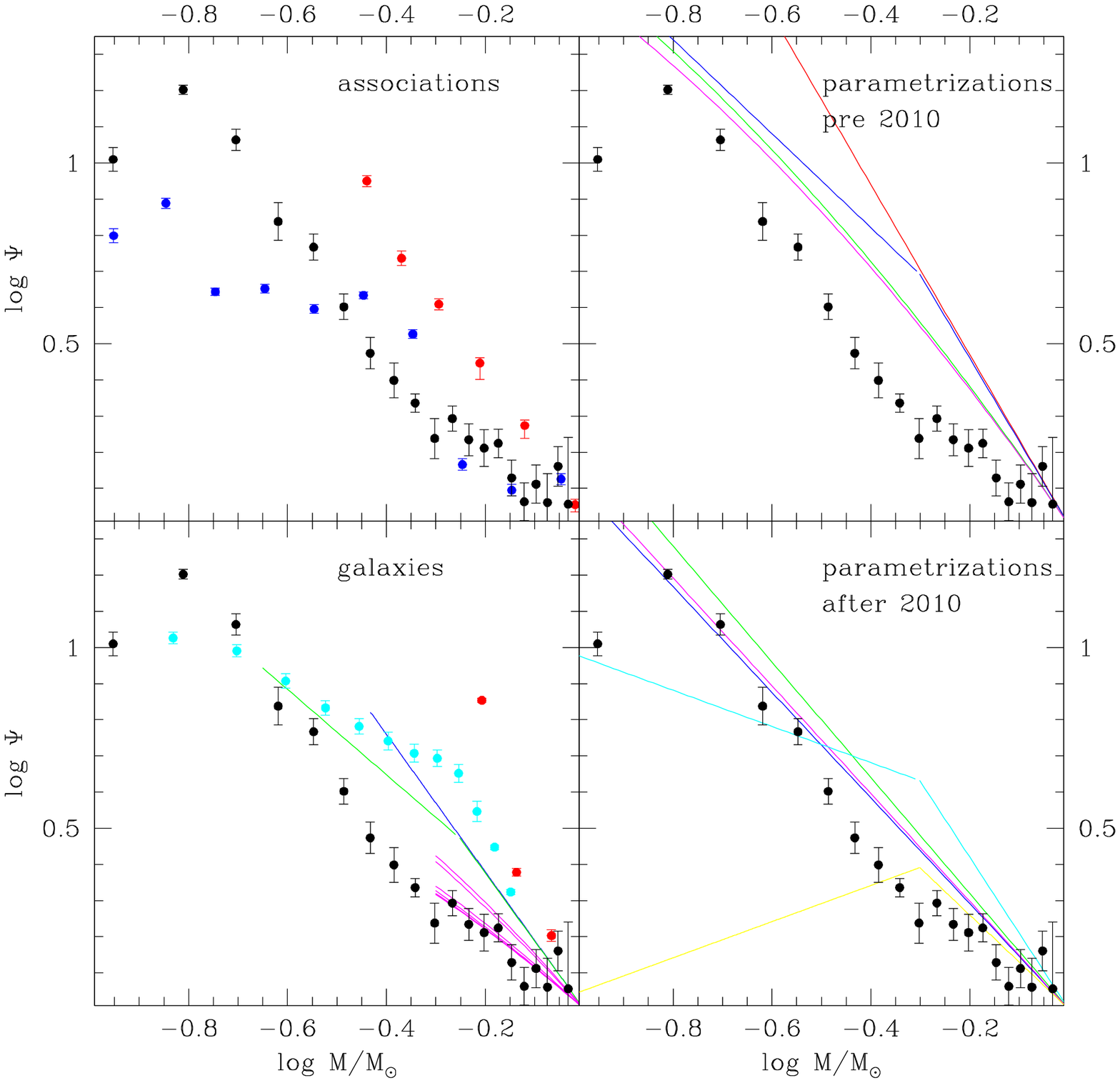}
 \caption{Comparison between the solar neighbourhood IMF estimated in this work
in the sub-solar mass regime (black dots)
 and the MFs estimated in associations (top-left panel: blue and red dots
 correspond to the Orion Nebula
 Cluster and LH95), galaxies (bottom-left panel: blue, red, green,
 magenta and cyan symbols correspond to the Large and Small
 Magellanic Cloud, Com Ber, various Ultra faint dwarfs and $\omega$ Cen,
 respectively) and the most popular parametric fit to the solar neighbourhood
 (top-right panel: red, blue, green and magenta lines correspond to the works by
 Salpeter 1955, Kroupa 2001, Chabrier 2003 and Miller \& Scalo 1979, respectively; 
 bottom-right panels: red, blue, green, magenta, cyan and yellow lines correspond to 
 the works by Dawson \& Schr{\"o}der 2010, Just \& Jahrei{\ss} 2010, Czekaj et al. 2014, Rybizki \& Just 2015, 
 Mor et al. 2018, 2019, respectively).
 All the MFs are normalized to their values at $1~M_{\odot}$.}
\label{comp}
\end{figure*}

\begin{figure*}
 \includegraphics[width=15cm]{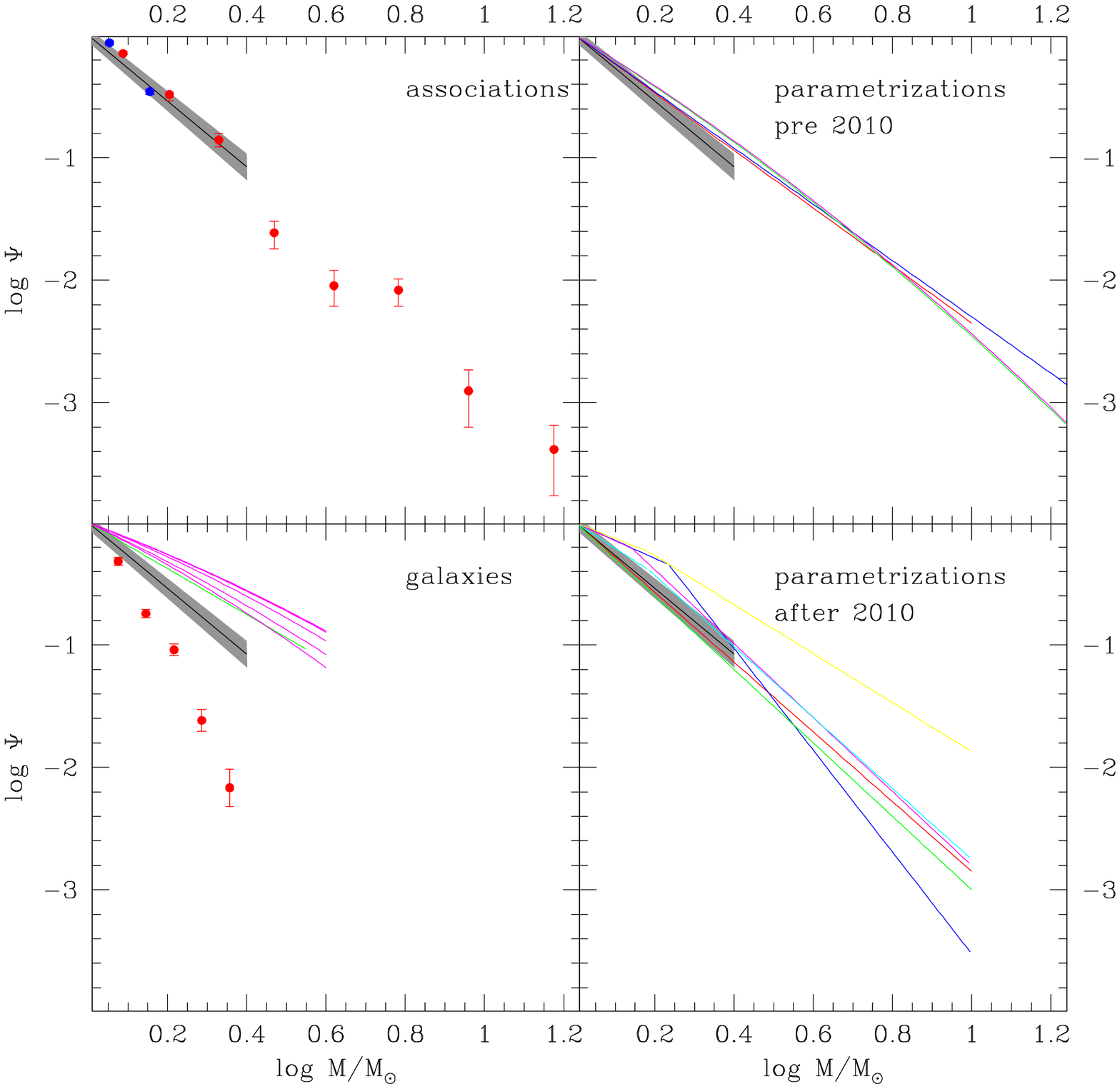}
 \caption{Same of Fig. \ref{comp} but for the super-solar mass regime. The slope
 derived in this work using the low-resolution SFH and its 1$\sigma$ uncertainty are represented by the solid line
 and the grey shaded area, respectively.}
\label{comp2}
\end{figure*}

In the top-right panels of Fig. \ref{comp} and \ref{comp2} the IMF derived in this work is
compared with those provided by
\citet{1955ApJ...121..161S}, \citet{1979ApJS...41..513M},
\citet{2001MNRAS.322..231K}, and \citet{2003PASP..115..763C}
in the two considered mass regimes, respectively.
The IMF estimated here is similar in the sub-solar regime to those of the considered works. 
Moreover, for the first time, a peak in the IMF at masses above the
hydrogen-burning limit is detected.
In the super-solar regime a good agreement is found when the high-resolution SFH 
is adopted, while in the low-resolution case the IMF estimated here is significantly 
steeper than that found by these authors. 

A better agreement with the IMF derived adopting the low-resolution SFH is instead found with the post-Hipparcos IMF determinations 
(bottom-right panels in Figs. \ref{comp} and \ref{comp2}).
\citet{2010MNRAS.404..917D} estimated a MF slope $\alpha=-2.85\pm0.15$ from a sample of Hipparcos 
stars with $M>0.9~M_{\odot}$ at heliocentric distances $<$100 pc and within 
$\pm$25 pc to the Galactic plane.
\citet{2010MNRAS.402..461J}
analysed the data from Hipparcos and the Catalog of Nearby Stars (at distances $<$200 pc) over a
mass range $0.5<M/M_{\odot}<10$. 
They found a broken power-law IMF with indexes
$\alpha=-1.46\pm0.1$ at $M<1.72~M_{\odot}$ and $\alpha=-4.16\pm0.12$ at larger
masses. Using the same dataset and a different prescription for the extinction
law, \citet{2015MNRAS.447.3880R} updated these values to $\alpha=-1.49\pm0.08$ at
$M<1.39~M_{\odot}$ and $\alpha=-3.02\pm0.06$ beyond this mass.
Similar results were obtained by \citet{2014A&A...564A.102C} who used 
the Galactic model by \citet{2003A&A...409..523R} to 
derive $\alpha\sim-3$ over a wide range of masses, although their sample is
limited at relatively bright magnitudes so that the constraint for stars 
with masses $M<1.5~M_{\odot}$ is less stringent.
\citet{2018A&A...620A..79M} performed a similar comparison on the same
dataset and
found $\alpha=-2.1_{-0.3}^{+0.1}$ in the mass range $0.5<M/M_{\odot}<1.53$ and
$\alpha=-2.9\pm0.2$ at larger masses ($\alpha=-3.7\pm0.2$ using a different
extinction map). Finally, \citet{2019A&A...624L...1M} fitted the Gaia DR2 data with the 
Besancon model and derived a slope $\alpha=-1.3\pm 0.3$ in the mass range 
$0.5<M/M_{\odot}<1.53$ and $\alpha=-1.9_{-0.2}^{+0.1}$ at 
larger masses. However, as in the present analysis, their MF slopes appear to strongly depend on the 
assumptions about the shape of the SFH: when they impose an exponential SFH, 
the high-mass slope becomes $\alpha=-2.5\pm 0.1$. 
At the low-mass extreme 
($M<0.5~M_{\odot}$) they found positive slopes $\alpha=+0.4_{-0.2}^{+0.6}$ and 
$\alpha=+0.5_{-0.8}^{+0.5}$ in the case of an exponential or of a non-parametric 
SFH, respectively. Given the very large uncertainties in this very-low mass range, 
the difference with respect to our work is not significant.

Considering that all the works quoted above adopt different
assumptions for the reddening, the Galactic structure and the adopted stellar
models, there is a surprisingly good agreement with the IMF estimated in this
work.

\subsection{Pleiades}
\label{comp_pleiades_sec}

\begin{figure*}
 \includegraphics[width=15cm]{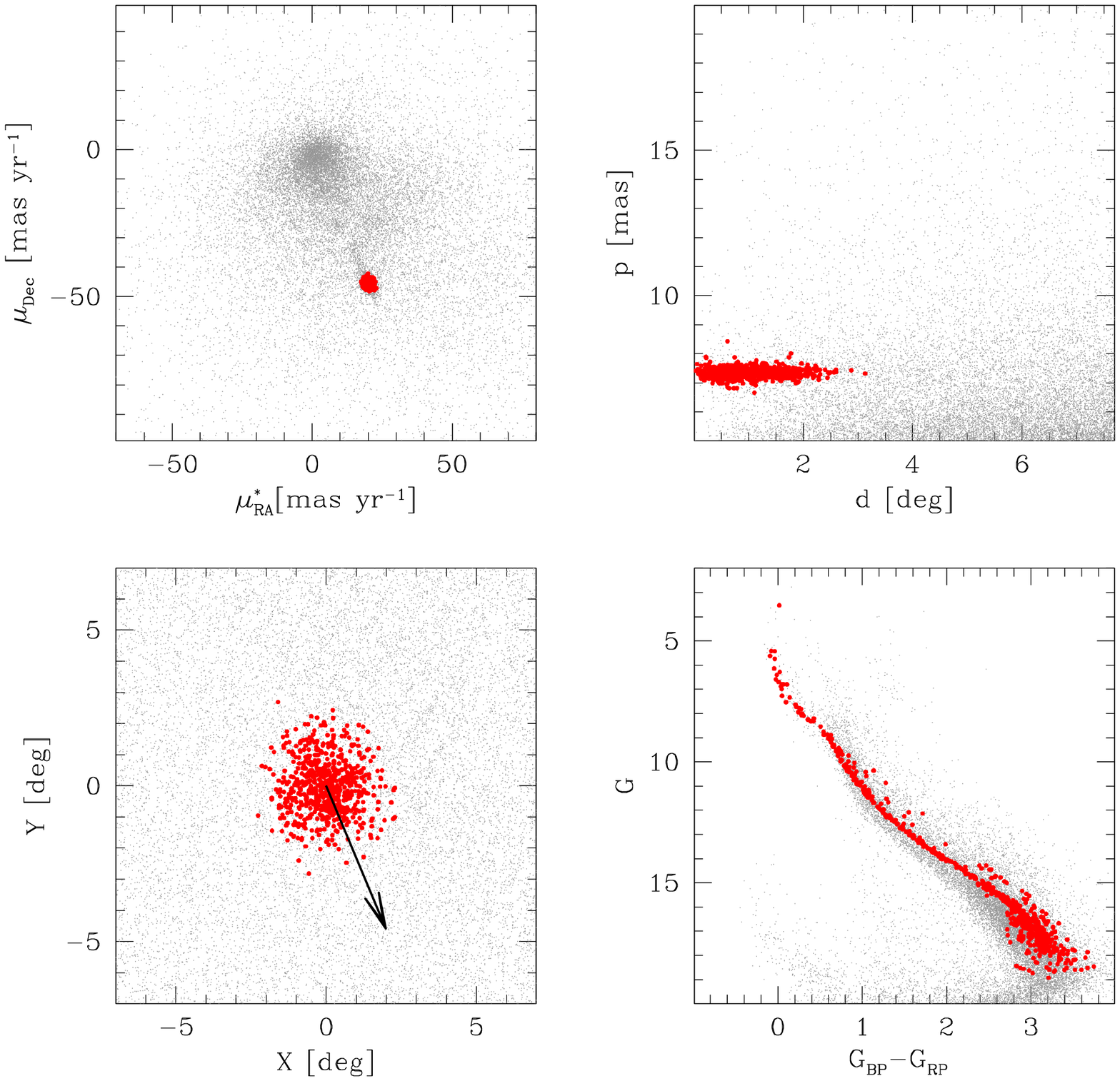}
 \caption{Distribution of stars with parallaxes $p>5~mas$ and distance from the
 Pleiades center $d<7.7^{\circ}$ in the proper motion (top-left panel),
 parallax (top-right panel), projected position (bottom-left) and
 colour-magnitude (bottom-right panel) diagrams. Bona-fide members are plotted
 with red dots. The direction of the bulk proper motion on the X-Y map is shown by an
 arrow.}
\label{pleisel}
\end{figure*}

\begin{figure}
 \includegraphics[width=8.6cm]{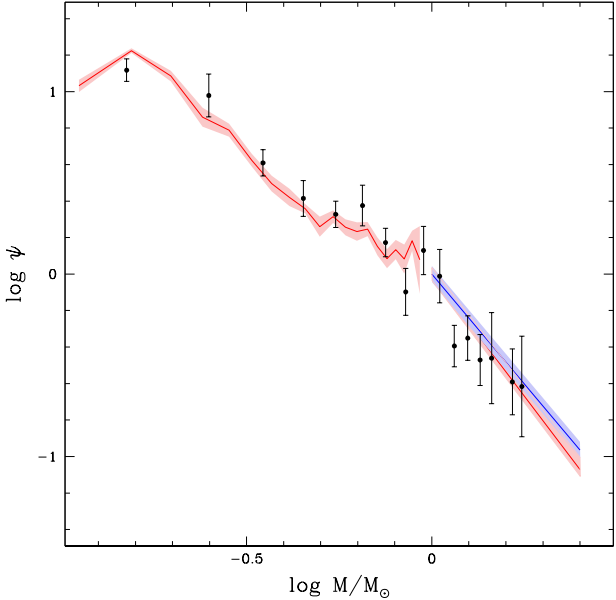}
 \caption{Comparison between the Pleiades MF (black dots) and that of the solar
 neighbourhood (red and blue lines correspond to the low- and high-resolution SFH case, respectively). 
 The shaded area represents the 1$\sigma$ uncertainty. All 
 MFs are normalized to their values at $1~M_{\odot}$.}
\label{pleia}
\end{figure}

\begin{figure*}
 \includegraphics[width=15cm]{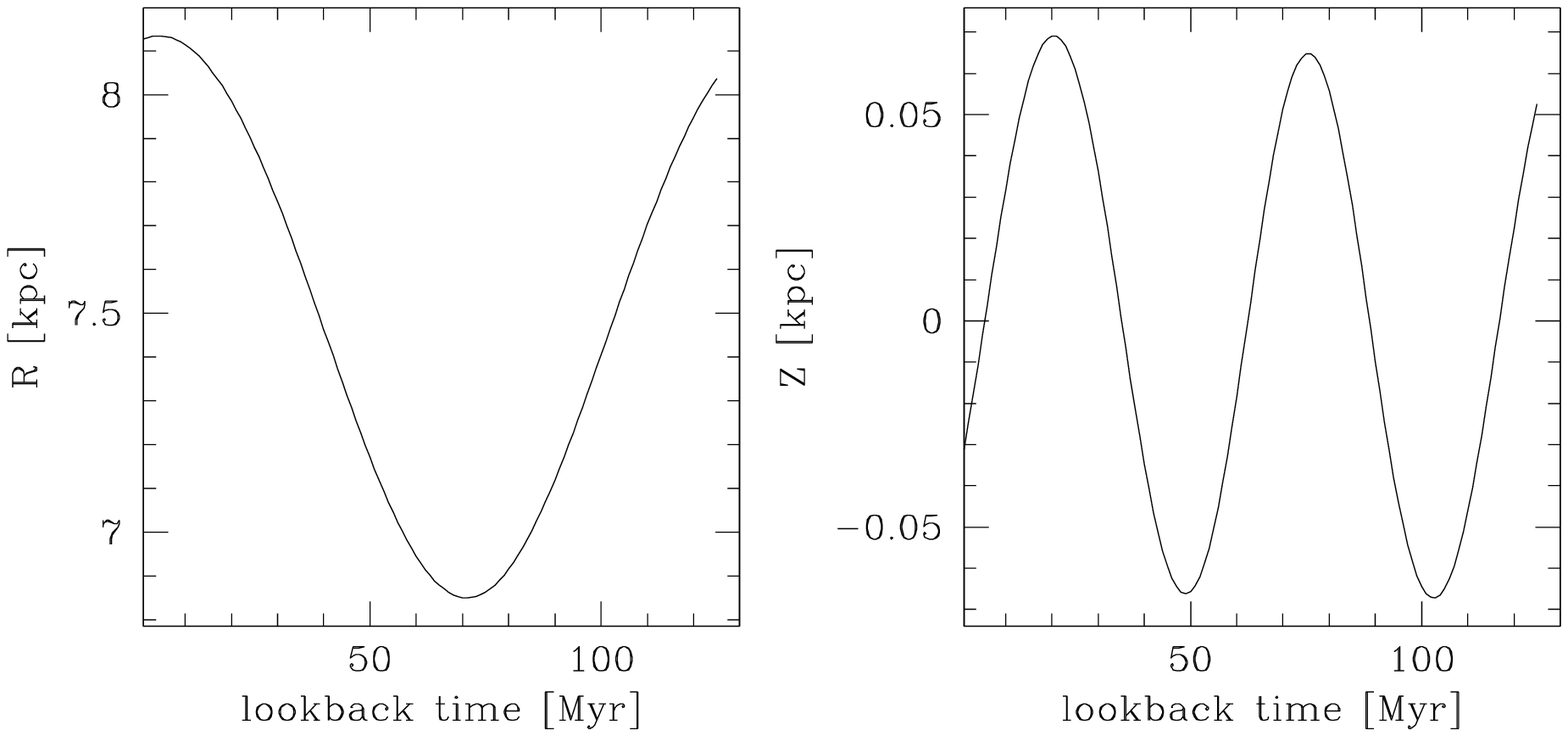}
 \caption{Reconstructed orbit of the Pleiades in the last 130 Myr. The variation
 of the projected distance on the Galactic plane and of the height above the
 Galactic plane as a function of the lookback time are shown in the left and
 right panels, respectively.}
\label{pleiorb}
\end{figure*}

As shown in Sect. \ref{methi_sec}, the procedure to derive the IMF in the solar
neighbourhood involves many free parameters and suffers from significant
systematic errors, in particular in the super-solar mass regime. While this is an unavoidable situation in the Galactic field,
a more robust estimate could be made in a nearby stellar system where stars are
all located at the same distance and have similar ages and chemical compositions
(see Sect. \ref{intro_sec}). Among the open clusters contained in the volume defined
for the {\it bright sample}, the Pleiades are young and massive enough to sample
a relatively wide range of masses with a good statistics.

In this cluster, Gaia proper motions and parallaxes allow to select member stars
with an unprecedented efficiency on the basis of the distribution of stars in
the 5D space formed by projected positions, proper motions and parallaxes.
In this space the Pleiades are clustered around a mean proper motion
$\langle \mu_{RA}^{*}\rangle=19.96\pm 0.04~mas/yr$, $\langle
\mu_{Dec}\rangle=-45.51\pm 0.04~mas/yr$ and a mean parallax
of $\langle p \rangle=7.335\pm 0.003~mas$ \citep[corresponding to a distance of
$136.34\pm 0.05$ pc in agreement with the interferometric distance estimated
by][]{2004Natur.427..326P}. The density of stars in this space was calculated
using a k-neighbour algorithm with k=10 and normalizing projected distances and
velocities to their r.m.s. Cluster members are then
defined as those objects lying in a region characterized by a density at
5$\sigma$ above the average background density calculated in a portion of this
space surrounding the region occupied by the bulk of cluster members.  
By using the above selection criterion I selected 674
bona-fide cluster members (see Fig. \ref{pleisel}).
To quantify the possible residual contamination from Galactic interlopers 
the same selection criterion was applied to a control field selected at the 
same Galactic latitude of the Pleiades and displaced by $\sim 13.5^{\circ}$ in 
longitude: only 1 star passed the above defined criterion indicating a contamination 
$<0.2\%$.
The maximum projected density in the cluster center is $<4\times
10^{-5}~arcsec^{-2}$, so crowding effects are negligible.
At the distance of the Pleiades, the completeness cuts of Gaia correspond to
masses of 0.13 $M_{\odot}$ and 2 $M_{\odot}$ i.e. comparable with those of the
Galactic field.

The Pleiades MF was derived in this mass range using the same technique described in Sect.
\ref{metlo_sec} assuming a single age ($log~t/yr=8.1$) and metallicity ([Fe/H=0])
derived from the comparison with MESA isochrones, in agreement with previous
literature determinations \citep{2009AJ....138.1292S,2018ApJ...863...67G}.
Absolute magnitudes were computed from eq. \ref{abs_eq} and assuming a 
reddening of E(B-V)=0.03 \citep{1986ApJ...309..311B}.
Different values of the binary fraction were tested by
randomly pairing a fraction of stars extracted from the MF. 
The best fit MF was chosen as the one providing the lowest value of 
the penalty function $\xi$
(eq. \ref{chi_eq}) and it is shown in Fig. \ref{pleia}.
It was obtained assuming a binary fraction of $f_{b}=37\%$, smaller 
than the 76\% estimated by \citet{2008ApJ...678..431C}.
The MF derived here is compatible with that estimated by the most comprehensive
studies conducted on this stellar system to date 
\citep{2003A&A...400..891M,2018A&A...617A..15O} in the high mass range, although
\citet{2018A&A...617A..15O} derive a significantly flatter MF with $\alpha=-1.12$ in 
the mass range $0.2<M/M_{\odot}<0.56$.

Qualitatively, the Pleiades MF is remarkably similar to that estimated in
the solar neighbourhood over its entire mass extent.
In particular, the MF steepens toward high masses with a possible break mass at
$M\sim 1~M_{\odot}$. A fit with a broken power law gives $\alpha=-1.6\pm 0.2$ and
$\alpha=-3.4\pm 0.3$ at masses below/above $1~M_{\odot}$. These slopes are slightly steeper 
but still compatible with those estimated in the Galactic field.
In particular, the Pleiades MF slope at masses $M>1~M_{\odot}$ is much more similar to that 
estimated in the solar neighbourhood when a low-resolution SFH is assumed.
Part of the difference could be due to the larger binary fraction estimated in
this cluster since, as shown in Fig. \ref{fbmf}, the binary fraction and the MF
slope in the low-mass range are correlated with large binary fractions
corresponding to steeper MF slopes. Given the large uncertainty associated to the
binary fraction estimate ($\epsilon_{f_{b}}\sim 10\%$), it is possible to obtain a good fit
to both the Pleiades and the solar neighbourhood assuming the same binary 
fraction and MF slope.

The MF of single and binary stars was integrated over the mass range
between the hydrogen-burning limit ($0.09~M_{\odot}$) and the mass at the tip of the Red Giant
Branch ($4.61~M_{\odot}$) predicted by the best fit MESA isochrone to estimate
the total cluster mass.
The contribution to the total mass of white dwarfs was estimated by integrating
from the tip of the Red Giant Branch to 8 $M_{\odot}$ and adopting the
initial-final mass relation of \citet{2008ApJ...676..594K}, and is found to be $\sim$1.2\%. 
The estimated total mass of the Pleiades is
$365\pm 15~M_{\odot}$. The radial cumulative mass distribution of member 
stars was also calculated and 
fitted with a \citet{1966AJ.....71...64K} model with central adimensional
parameter $W_{0}=3$ and a half-mass radius of $r_{h}=1.59^{\circ}$
corresponding to 3.8 pc at the distance of the Pleiades.
The mass, half-mass radius and number of objects estimated above were used
to compute the half-mass relaxation time \citep{1987degc.book.....S} of the Pleiades 
$t_{rh}=127~Myr$. This timescale is comparable to the age of this stellar system
so that, while the effect of two-body relaxation could be measurable in massive
stars, it could not have significantly altered the global shape of the 
IMF \citep{2003MNRAS.340..227B}. 

By converting the mean proper motions to projected velocities and adopting a
radial velocity of 3.503 km/s \citep{2014A&A...562A..54C}, I reconstructed the orbit of the
Pleiades in the last 130 Myr in a \citet{1995ApJ...451..598J} potential using a fourth-order
Runge-Kutta integrator. The orbit of the Pleiades is confined to a small region
of the R-Z plane oscillating between $6.8<R/kpc<8.1$ and reaching a maximum
height above the Galactic plane of 70 pc (see Fig. \ref{pleiorb}). 

Summarizing, being dynamically young and having been likely formed in the solar
neighbourhood, the Pleiades are a representative episode of recent star 
formation in the solar vicinity whose IMF has still not been affected by
dynamical evolution. The similarity between their MF and that estimated in the
field supports the robustness of the IMF derived in Sect. \ref{reshi_sec}.

\subsection{Comparison with IMF of dynamically unrelaxed stellar systems}
\label{comp_other_sec}

In the left panels of Fig.s \ref{comp} and \ref{comp2}, the IMF is compared with those available for a sample of 
dynamically unrelaxed stellar systems i.e. associations and galaxies.
Among nearby associations I considered the deep MF estimates for the Orion
Nebula Cluster \citep[with {[Fe/H]} $\sim$0 and an average age of 2.3 
Myr; ][]{2012ApJ...748...14D} and the LH95 in the
Large Magellanic Cloud \citep[{[Fe/H]} $\sim$ -0.3, t$\sim$ 4 
Myr; ][]{2009ApJ...696..528D}. In the sub-solar mass regime
the IMF estimated in this work lies between those estimated for these two
objects, while at large masses a good agreement is found with the LH95 MF.
Note that, according to the quoted errors, at $M<0.4~M_{\odot}$ the MFs of these two associations
differ significantly. To quantify this difference, a $\chi^{2}$ test was performed 
selecting the portion of the MFs of these two associations ($log~N_{a}$ and $log~N_{b}$, respectively) 
in the n bins in the common mass range. 
A normalization factor $c$, needed to account for the different mass of the two associations, 
was calculated as that providing the best match between the two MFs (i.e. the 
inverse-variance weighted mean of the MF differences in the mass range considered) 
and applied. 
The quantity 
\begin{equation*}
\chi^{2}=\sum_{i=1}^{n} (log N_{a,i}+c-log N_{b,i})/\sqrt{\epsilon_{a,i}^{2}+\epsilon_{b,i}^{2}}
\end{equation*}
was then calculated. If the two estimated MFs are extracted from the same parent distribution, 
the function above should be distributed as a $\chi^{2}$ with n-1 
degrees of freedom.
The associated probability ($P_{\chi^{2}}$) is therefore an indicator of the similarity of the two MFs.
The above test gives a probability 
$<10^{-3}$ that the MFs of the two associations
are extracted from the same parent distribution. 
These systems were analysed by the same group with the same technique, so 
it is unlikely that the difference above can be attributed to systematic errors. 
Moreover, given their young ages, this 
difference cannot be interpreted as a result of any dynamical process occurring 
on such a short timescale and could therefore be primordial.

I also considered the MF measured in a sample of nearby galaxies: the Large and
Small Magellanic Clouds \citep{2006ApJ...641..838G,2013ApJ...763..110K}, the
sample of Ultra faint dwarfs by \citet{2018ApJ...855...20G,2018ApJ...863...38G}
and $\omega$ Centauri \citep{2007MNRAS.381.1575S}, the massive globular cluster 
with a half-mass relaxation time longer than its age and supposed to be the
remnant of an accreted galaxy because of its large metallicity spread \citep{2002ARA&A..40..487F}. 
The galaxies considered span a wide range in metallicity: from $[Fe/H]\sim-2.7$ 
for the Ultra-faint dwarfs \citep{2015ApJ...808...95S} to [Fe/H]=-0.3 for the Large Magellanic Clouds \citep{1998AJ....115..605L}.
Also in this case, at low-masses there is a large spread in the MF measured in 
these objects. The solar neighbourhood IMF estimated here stands at the lower 
boundary of the distribution of the considered MFs, being significantly flatter
than the steepest ones such as e.g. that of the Small Magellanic Cloud. At masses
above the solar mass, for both of the two assumptions on the SFH, the IMF 
estimated here is comprised between those of the
Ultra faint dwarfs and that of the Small Magellanic Clouds. Again, in spite of
the large uncertainties involved, there is a wide spread among the various
galaxies. 

To test the 
hypothesis that the differences between the MF of these stellar systems and that 
estimated for the solar neighbourhood are due to random uncertainties, a 
$\chi^{2}$ test was performed (see above). Significant ($P_{\chi^{2}}<0.3\%$) 
differences were found with respect to the Orion Nebula Cluster, the Small 
Magellanic Clouds and $\omega$ Cen, while for the other systems the differences 
are significant at values $0.3\%<P_{\chi^{2}}<5\%$. The same result is 
obtained when adopting the solar neighbourhood MF estimated in the low-mass range using 
the PARSEC isochrones and the MF derived adopting either the low- or the 
high-resolution SFH in the super-solar regime.

\section{summary}
\label{summ_sec}

I used the most complete and accurate data set available to date provided by the
2nd data release of the Gaia mission to derive the IMF of the solar
neighbourhood. The resulting IMF is well represented by a segmented power-law 
with two breaks at characteristic masses. 

The first break occurs in the very low-mass regime where
the IMF clearly flattens (and possibly decreases) at $M< 0.15~M_{\odot}$. This feature does not depend on the adopted stellar
models, fraction of binaries or sample completeness.  
Unfortunately, because of the uncertainties on the mass-luminosity relation at very-low masses, 
it is not clear whether the deficiency of stars observed at masses below such a 
break is significant.
A similar claim was made by \citet{1997ApJ...476L..19D} on the basis of
the analysis of the MF of a sample of Galactic globular clusters. That evidence was
however questioned because of the uncertain completeness of their data at faint
magnitudes and the possible occurrence of dynamical effects in these old stellar
systems \citep{1999A&A...345..485P}.

The existence of a peak in the IMF is predicted by star formation theories
although it is not clear if its position should lie in the stellar or sub-stellar regime.
It is interesting to analyse the observational evidence presented here in the
light of the two main star formation theories.
Theories based on the fragmentation on Jeans mass
scale + accretion \citep{1992MNRAS.256..641L,2008ApJ...684..395H} predict a peak mass 
corresponding to the smallest self-gravitating mass in a cloud able
to collapse. This characteristic mass depends on the ratio between the thermal 
Jeans mass and the square of the cloud Mach number \citep{2014ApJ...796...75C}. 
These quantities are functions of the
thermodynamical properties of the original cloud (temperature, mean molecular
weight, density) and on the relative efficiency of those processes affecting 
turbulent and magnetic pressure. Clouds below this critical mass should not begin star formation
unless a local temperature/density/pressure fluctuation causes a decrease of the
ratio mentioned above. In this picture, the evidence shown here suggests that in the solar neighbourhood
star formation occurred in conditions (low Mach number, low molecular weight,
high temperature, small cloud size) favouring the emergence of a peak mass at
a relatively high mass.
On the other hand, in theories based on the accretion/feedback balance 
\citep{1996ApJ...464..256A} the peak in the MF is given by the
minimum sound speed (i.e. its thermal value at the cloud temperature) while
stars below this mass can still form as a result of fluctuations in the other
involved parameters. From their calculation, the IMF peak should lie at
$M\sim 0.07 M_{\odot}$ slightly lower than the value determined here.
Also in this scenario, the existence of a peak in the IMF at $M\sim 0.15
M_{\odot}$ suggests conditions leading to a larger minimum sound speed (i.e. larger temperature, lower
molecular weight). 

In the mass range $0.15<M/M_{\odot}<1$ the MF is well represented by a single
power-law with mean slope $\alpha=-1.34\pm0.07$. This mean value is in agreement within the
uncertainties with the average slopes in the same mass range found in all 
previous works
\citep{1979ApJS...41..513M,2001MNRAS.322..231K,2003ApJ...586L.133C,2010MNRAS.402..461J,2015MNRAS.447.3880R}
although some of these last works adopted different functional forms for the MF.
Unfortunately, it is
not an easy task to distinguish among the various analytical representations
(broken power-law, log-normal, tampered log-normal, etc.) because they almost overlap in this mass range. Moreover, 
systematic uncertainties in the mass-luminosity relation can create artifacts 
altering the shape of the MF at small scales (see Sect.\ref{reslo_sec}).
I do not notice any change of slope at $0.5~M_{\odot}$ as reported by \citet{2001MNRAS.322..231K}, in agreement with
the works by
\citet{1979ApJS...41..513M}, \citet{2003ApJ...586L.133C} and \citet{2015MNRAS.447.3880R}.
While the uncertainties in the mass-luminosity relation described above can hide
the evidence of such a break, a significant slope change in this intermediate
mass range $0.15<M/M_{\odot}<1$ is not supported by the data analysed here. 

At masses larger than 1 $M_{\odot}$, if a smoothly varying SFH is assumed, the average IMF slope is found to be
$\alpha=-2.68\pm 0.09$, significantly steeper than those found in works
\citep{1955ApJ...121..161S,1979ApJS...41..513M,2001MNRAS.322..231K} dated 
before the most extensive astrometric missions \citep[Hipparcos and 
Gaia; although a similar value was reported by][]{1993MNRAS.262..545K} and compatible with those found in subsequent analyses
\citep{2010MNRAS.404..917D,2010MNRAS.402..461J,2015MNRAS.447.3880R,2019A&A...624L...1M}.
Given the improvement in sample size and accuracy of these surveys, these last
results seem to be more robust.
Unfortunately, this portion of the MF is subject to many systematic uncertainties
linked to the modelling of the age/metallicity/distance/reddening variations and on the 
uncertainties in the SFH (see Sect. \ref{reshi_sec}). Indeed, a flatter IMF 
slope would be compatible with the data if a bursty SFH characterized by rapid variations 
of the star formation rate were adopted.

A steep IMF is also suggested by the IMF measured in the Pleiades which formed in a single burst of star 
formation and where no significant spread in metallicity/reddening/distance is 
expected.
The analysis performed in Sect. \ref{comp_pleiades_sec} shows 
that this cluster formed in the solar vicinity and should not have experienced 
significant dynamical evolution. In the commonly accepted scenario, where the
Galactic field population originates in clusters and associations which 
dissolve in a quick timescale \citep{2003ApJ...598.1076K,2012MNRAS.426.3008K,2018A&A...620A..39J}, 
the IMF measured in the solar neighbourhood is 
therefore not representative of a single star formation event but is the 
superposition of contributions of many small episodes.
The Pleiades are thus an example in which such
building blocks retain the information on the IMF in their PDMF. 
However, it must be considered that low-mass stars move away from their original
site of birth more efficiently than massive ones because of the velocity drift induced by primordial mass 
segregation and competitive accretion
\citep{1997MNRAS.285..201B,2007ApJ...655L..45M} and because of their long
lifetime, have more time to distribute over the Galactic plane.
Therefore, samples of stars covering a wide portion of the field contain preferentially
low-mass stars while high-mass ones are confined in more compact portions of the
phase-space close to the position of their original birth sites. 
This effect is responsible for a large cosmic variance in the Galactic disk IMF, 
with a bias toward measuring steeper IMF slope in the field \citep{2018MNRAS.480.2449P}.

The slopes of the IMF at intermediate and high-masses have been interpreted in
different ways by different star formation theories. 
According to theories of accretion onto Jeans mass scale fragments, accretion
occurs in a competitive way at different characteristic stellar radii in
different cluster regions according to the 
relative contribution to the overall cluster potential of gas and stars. In this model, high-mass 
stars preferentially form in a clustered environment in the 
central region of the proto-clusters and accrete the majority of their mass at the 
Bondi-Hoyle radius. 
The strong mass dependence of the accretion rate results in a steeper IMF with respect 
to that predicted by simple tidal accretion onto individual fragments typical of low-mass stars. 
The transition between the accretion-
and fragmentation-dominated regimes depends on the prescriptions adopted for
primordial mass segregation and on the relative distribution of gas and stars 
at early stages.
In the two regimes considered above the IMF should be characterized by an asymptotic 
slope $\alpha\sim -1.5$ at low-masses and $\alpha\sim -2.5$ at high-masses \citep{2001MNRAS.324..573B}.
These values are in agreement with those found in the present analysis.
In this picture, the results presented here support a transition mass close to 1 $M_{\odot}$
dividing the mass spectrum in two ranges characterized by different slopes.
Alternatively, in the \citet{1996ApJ...464..256A} theory, the change of IMF slope 
is due to the different
mass-luminosity relation of young stellar objects as a function of their mass
\citep{1996ApJ...464..256A}. 
Indeed, while the luminosity of low mass objects is
determined by the infall rate, massive ones generate a significant
fraction of luminosity through gravitational contraction, deuterium- and
eventually hydrogen-burning. In this case, the transition is extremely smooth
and should occur at $\sim 3.1 M_{\odot}$.
Note that in this simplified model, even assuming a dependence of the IMF
slope on the sound speed distribution alone (i.e. the parameter with the largest
contribution to the final stellar mass), the IMF slope 
should have a small variation $-2.1<\alpha<-1.7$ across the entire mass 
range, much smaller than that observed in this analysis. The stochastic 
variation of the other parameters involved smoothes the overall IMF further,
reducing the slope variation and
enhancing the tension with observations. However, the models above contain
many simplified recipes to model the complex set of physical processes at work
in star forming regions so that it is hard to rule out this scenario on the
basis of relatively small differences in the shape of the IMF.

The comparison of the IMF measured in the solar neighbourhood with those
estimated in other non-collisional environments (galaxies and associations)
reveals a significant degree of variability incompatible with the quoted
uncertainties. This would imply that the IMF is not Universal.
However, there is no clear trend of the IMF slope with either metallicity or
environment. 
Consider that, as shown in Sect. \ref{reshi_sec}, systematic effects can alter
the MF slope by a large amount. Since the MFs considered have been estimated by
different groups adopting different prescriptions, it is possible that any
existing trend could have been erased by such systematic 
errors.

The analysis presented here will further benefit from the next Gaia data releases.
Indeed, besides the incremental improvement of the photometric and astrometric 
performances, the completeness at bright magnitudes should be established allowing the
use of the brightest portion of the luminosity function to constrain the IMF at
its high-mass end ($M>2.5~M_{\odot}$) and to constrain the SFH at recent
epochs with better resolution. Moreover, starting from DR3 metallicities, 
reddening and a classification of 
binaries will be provided, thus allowing to calibrate the model
parameters better, accounting for their variation across the disc and to replace the
statistical population synthesis approach with a 
star-by-star modelling.

\section*{Acknowledgments}
I warmly thank Michele Bellazzini, Paola Marrese and Michele Cignoni for 
useful discussions and Giovanna Stirpe for the careful reading of the manuscript. I also thank the anonymous referee for his/her helpful 
comments and suggestions.

\label{lastpage}


\begin{thebibliography}{99}

\bibitem[\protect\citeauthoryear{Adams \& Fatuzzo}{1996}]{1996ApJ...464..256A} Adams F.~C., Fatuzzo M., 1996, ApJ, 464, 256 
\bibitem[\protect\citeauthoryear{Adams \& Laughlin}{1996}]{1996ApJ...468..586A} Adams F.~C., Laughlin G., 1996, ApJ, 468, 586 
\bibitem[\protect\citeauthoryear{Arenou et al.}{2018}]{2018A&A...616A..17A} Arenou F., et al., 2018, A\&A, 616, A17 
\bibitem[\protect\citeauthoryear{Ballero, Kroupa, \& Matteucci}{2007}]{2007A&A...467..117B} Ballero S.~K., Kroupa P., Matteucci F., 2007, A\&A, 467, 117 
\bibitem[\protect\citeauthoryear{Bastian, Covey, \& Meyer}{2010}]{2010ARA&A..48..339B} Bastian N., Covey K.~R., Meyer M.~R., 2010, ARA\&A, 48, 339 
\bibitem[\protect\citeauthoryear{Baumgardt \& Makino}{2003}]{2003MNRAS.340..227B} Baumgardt H., Makino J., 2003, MNRAS, 340, 227 
\bibitem[\protect\citeauthoryear{Bernard}{2018}]{2018IAUS..330..148B} Bernard E.~J., 2018, IAUS, 330, 148 
\bibitem[\protect\citeauthoryear{Bellazzini, Ferraro, \& Pancino}{2001}]{2001ApJ...556..635B} Bellazzini M., Ferraro F.~R., Pancino E., 2001, ApJ, 556, 635 
\bibitem[\protect\citeauthoryear{Bonnell et al.}{1997}]{1997MNRAS.285..201B} Bonnell I.~A., Bate M.~R., Clarke C.~J., Pringle J.~E., 1997, MNRAS, 285, 201 
\bibitem[\protect\citeauthoryear{Bonnell et al.}{2001}]{2001MNRAS.324..573B} Bonnell I.~A., Clarke C.~J., Bate M.~R., Pringle J.~E., 2001, MNRAS, 324, 573 
\bibitem[\protect\citeauthoryear{Bonnell, Clarke, \& Bate}{2006}]{2006MNRAS.368.1296B} Bonnell I.~A., Clarke C.~J., Bate M.~R., 2006, MNRAS, 368, 1296 
\bibitem[\protect\citeauthoryear{Bovy}{2017}]{2017MNRAS.470.1360B} Bovy J., 2017, MNRAS, 470, 1360 
\bibitem[\protect\citeauthoryear{Breger}{1986}]{1986ApJ...309..311B} Breger M., 1986, ApJ, 309, 311 
\bibitem[\protect\citeauthoryear{Bressan et al.}{2012}]{2012MNRAS.427..127B} Bressan A., Marigo P., Girardi L., Salasnich B., Dal Cero C., Rubele S., Nanni A., 2012, MNRAS, 427, 127 
\bibitem[\protect\citeauthoryear{Casagrande \& VandenBerg}{2018}]{2018MNRAS.479L.102C} Casagrande L., VandenBerg D.~A., 2018, MNRAS, 479, L102 
\bibitem[\protect\citeauthoryear{Chabrier}{2001}]{2001ApJ...554.1274C} Chabrier G., 2001, ApJ, 554, 1274 
\bibitem[\protect\citeauthoryear{Chabrier}{2003a}]{2003ApJ...586L.133C} Chabrier G., 2003a, ApJ, 586, L133 
\bibitem[\protect\citeauthoryear{Chabrier}{2003b}]{2003PASP..115..763C} Chabrier G., 2003b, PASP, 115, 763 
\bibitem[\protect\citeauthoryear{Chabrier, Hennebelle, \& Charlot}{2014}]{2014ApJ...796...75C} Chabrier G., Hennebelle P., Charlot S., 2014, ApJ, 796, 75 
\bibitem[\protect\citeauthoryear{Choi et al.}{2016}]{2016ApJ...823..102C} Choi J., Dotter A., Conroy C., Cantiello M., Paxton B., Johnson B.~D., 2016, ApJ, 823, 102 
\bibitem[\protect\citeauthoryear{Conrad et al.}{2014}]{2014A&A...562A..54C} Conrad C., et al., 2014, A\&A, 562, A54 
\bibitem[\protect\citeauthoryear{Converse \& Stahler}{2008}]{2008ApJ...678..431C} Converse J.~M., Stahler S.~W., 2008, ApJ, 678, 431 
\bibitem[\protect\citeauthoryear{Courteau et al.}{2014}]{2014RvMP...86...47C} Courteau S., et al., 2014, RvMP, 86, 47 
\bibitem[\protect\citeauthoryear{Czekaj et al.}{2014}]{2014A&A...564A.102C} Czekaj M.~A., Robin A.~C., Figueras F., Luri X., Haywood M., 2014, A\&A, 564, A102 
\bibitem[\protect\citeauthoryear{Da Rio, Gouliermis, \& Henning}{2009}]{2009ApJ...696..528D} Da Rio N., Gouliermis D.~A., Henning T., 2009, ApJ, 696, 528 
\bibitem[\protect\citeauthoryear{Da Rio et al.}{2012}]{2012ApJ...748...14D} Da Rio N., Robberto M., Hillenbrand L.~A., Henning T., Stassun K.~G., 2012, ApJ, 748, 14 
\bibitem[\protect\citeauthoryear{Dawson \& Schr{\"o}der}{2010}]{2010MNRAS.404..917D} Dawson S.~A., Schr{\"o}der K.-P., 2010, MNRAS, 404, 917 
\bibitem[\protect\citeauthoryear{De Marchi \& Paresce}{1997}]{1997ApJ...476L..19D} De Marchi G., Paresce F., 1997, ApJ, 476, L19 
\bibitem[\protect\citeauthoryear{Demarque \& Mengel}{1973}]{1973A&A....22..121D} Demarque P., Mengel J.~G., 1973, A\&A, 22, 121 
\bibitem[\protect\citeauthoryear{Dib}{2014}]{2014MNRAS.444.1957D} Dib S., 2014, MNRAS, 444, 1957 
\bibitem[\protect\citeauthoryear{Drimmel \& Spergel}{2001}]{2001ApJ...556..181D} Drimmel R., Spergel D.~N., 2001, ApJ, 556, 181 
\bibitem[\protect\citeauthoryear{Duquennoy \& Mayor}{1991}]{1991A&A...248..485D} Duquennoy A., Mayor M., 1991, A\&A, 248, 485 
\bibitem[\protect\citeauthoryear{El-Badry, Weisz, \& Quataert}{2017}]{2017MNRAS.468..319E} El-Badry K., Weisz D.~R., Quataert E., 2017, MNRAS, 468, 319 
\bibitem[\protect\citeauthoryear{ESA}{1997}]{1997ESASP1200.....E} ESA, 1997, ESASP, 1200,  
\bibitem[\protect\citeauthoryear{Flewelling}{2018}]{2018AAS...23143601F} Flewelling H., 2018, AAS, 231, 436.01 
\bibitem[\protect\citeauthoryear{Freeman \& Bland-Hawthorn}{2002}]{2002ARA&A..40..487F} Freeman K., Bland-Hawthorn J., 2002, ARA\&A, 40, 487 
\bibitem[\protect\citeauthoryear{Gaia Collaboration et al.}{2018}]{2018A&A...616A...1G} Gaia Collaboration, et al., 2018, A\&A, 616, A1 
\bibitem[\protect\citeauthoryear{Geha et al.}{2013}]{2013ApJ...771...29G} Geha M., et al., 2013, ApJ, 771, 29 
\bibitem[\protect\citeauthoryear{Gennaro et al.}{2018a}]{2018ApJ...855...20G} Gennaro M., et al., 2018a, ApJ, 855, 20 
\bibitem[\protect\citeauthoryear{Gennaro et al.}{2018b}]{2018ApJ...863...38G} Gennaro M., et al., 2018b, ApJ, 863, 38 
\bibitem[\protect\citeauthoryear{Gossage et al.}{2018}]{2018ApJ...863...67G} Gossage S., Conroy C., Dotter A., Choi J., Rosenfield P., Cargile P., Dolphin A., 2018, ApJ, 863, 67 
\bibitem[\protect\citeauthoryear{Gouliermis, Brandner, \& Henning}{2006}]{2006ApJ...641..838G} Gouliermis D., Brandner W., Henning T., 2006, ApJ, 641, 838 
\bibitem[\protect\citeauthoryear{Hennebelle}{2012}]{2012A&A...545A.147H} Hennebelle P., 2012, A\&A, 545, A147 
\bibitem[\protect\citeauthoryear{Hennebelle \& Chabrier}{2008}]{2008ApJ...684..395H} Hennebelle P., Chabrier G., 2008, ApJ, 684, 395 
\bibitem[\protect\citeauthoryear{Hoversten \& Glazebrook}{2008}]{2008ApJ...675..163H} Hoversten E.~A., Glazebrook K., 2008, ApJ, 675, 163 
\bibitem[\protect\citeauthoryear{Jappsen et al.}{2005}]{2005A&A...435..611J} Jappsen A.-K., Klessen R.~S., Larson R.~B., Li Y., Mac Low M.-M., 2005, A\&A, 435, 611 
\bibitem[\protect\citeauthoryear{Je{\v r}{\'a}bkov{\'a} et al.}{2018}]{2018A&A...620A..39J} Je{\v r}{\'a}bkov{\'a} T., Hasani Zonoozi A., Kroupa P., Beccari G., Yan Z., Vazdekis A., Zhang Z.-Y., 2018, A\&A, 620, A39 
\bibitem[\protect\citeauthoryear{Johnston, Spergel, \& Hernquist}{1995}]{1995ApJ...451..598J} Johnston K.~V., Spergel D.~N., Hernquist L., 1995, ApJ, 451, 598 
\bibitem[\protect\citeauthoryear{Just \& Jahrei{\ss}}{2010}]{2010MNRAS.402..461J} Just A., Jahrei{\ss} H., 2010, MNRAS, 402, 461 
\bibitem[\protect\citeauthoryear{Kalirai et al.}{2008}]{2008ApJ...676..594K} Kalirai J.~S., Hansen B.~M.~S., Kelson D.~D., Reitzel D.~B., Rich R.~M., Richer H.~B., 2008, ApJ, 676, 594 
\bibitem[\protect\citeauthoryear{Kalirai et al.}{2013}]{2013ApJ...763..110K} Kalirai J.~S., et al., 2013, ApJ, 763, 110 
\bibitem[\protect\citeauthoryear{Karim \& Mamajek}{2017}]{2017MNRAS.465..472K} Karim M.~T., Mamajek E.~E., 2017, MNRAS, 465, 472 
\bibitem[\protect\citeauthoryear{King}{1966}]{1966AJ.....71...64K} King I.~R., 1966, AJ, 71, 64 
\bibitem[\protect\citeauthoryear{Kroupa}{1995}]{1995ApJ...453..358K} Kroupa P., 1995, ApJ, 453,  
\bibitem[\protect\citeauthoryear{Kroupa}{2001}]{2001MNRAS.322..231K} Kroupa P., 2001, MNRAS, 322, 231 
\bibitem[\protect\citeauthoryear{Kroupa}{2002}]{2002Sci...295...82K} Kroupa P., 2002, Sci, 295, 82 
\bibitem[\protect\citeauthoryear{Kroupa \& Weidner}{2003}]{2003ApJ...598.1076K} Kroupa P., Weidner C., 2003, ApJ, 598, 1076
\bibitem[\protect\citeauthoryear{Kroupa, Tout, \& Gilmore}{1991}]{1991MNRAS.251..293K} Kroupa P., Tout C.~A., Gilmore G., 1991, MNRAS, 251, 293 
\bibitem[\protect\citeauthoryear{Kroupa, Tout, \& Gilmore}{1993}]{1993MNRAS.262..545K} Kroupa P., Tout C.~A., Gilmore G., 1993, MNRAS, 262, 545 
\bibitem[\protect\citeauthoryear{Kruijssen}{2012}]{2012MNRAS.426.3008K} Kruijssen J.~M.~D., 2012, MNRAS, 426, 3008 
\bibitem[\protect\citeauthoryear{Lacy}{1978}]{1978ApJ...226..138L} Lacy C.~H., 1978, ApJ, 226, 138 
\bibitem[\protect\citeauthoryear{Lallement et al.}{2019}]{2019A&A...625A.135L} Lallement R., et al., 2019, A\&A, 625, A135
\bibitem[\protect\citeauthoryear{Lamers, Baumgardt, \& Gieles}{2013}]{2013MNRAS.433.1378L} Lamers H.~J.~G.~L.~M., Baumgardt H., Gieles M., 2013, MNRAS, 433, 1378 
\bibitem[\protect\citeauthoryear{Larson}{1978}]{1978MNRAS.184...69L} Larson R.~B., 1978, MNRAS, 184, 69 
\bibitem[\protect\citeauthoryear{Larson}{1992}]{1992MNRAS.256..641L} Larson R.~B., 1992, MNRAS, 256, 641 
\bibitem[\protect\citeauthoryear{Larson}{1998}]{1998MNRAS.301..569L} Larson R.~B., 1998, MNRAS, 301, 569 
\bibitem[\protect\citeauthoryear{Limber}{1960}]{1960ApJ...131..168L} Limber D.~N., 1960, ApJ, 131, 168 
\bibitem[\protect\citeauthoryear{Lindegren et al.}{2018}]{2018A&A...616A...2L} Lindegren L., et al., 2018, A\&A, 616, A2 
\bibitem[\protect\citeauthoryear{Luck et al.}{1998}]{1998AJ....115..605L} Luck R.~E., Moffett T.~J., Barnes T.~G., III, Gieren W.~P., 1998, AJ, 115, 605 
\bibitem[\protect\citeauthoryear{Luri et al.}{2018}]{2018A&A...616A...9L} Luri X., et al., 2018, A\&A, 616, A9 
\bibitem[\protect\citeauthoryear{Marrese et al.}{2019}]{2019A&A...621A.144M} Marrese P.~M., Marinoni S., Fabrizio M., Altavilla G., 2019, A\&A, 621, A144 
\bibitem[\protect\citeauthoryear{McMillan, Vesperini, \& Portegies Zwart}{2007}]{2007ApJ...655L..45M} McMillan S.~L.~W., Vesperini E., Portegies Zwart S.~F., 2007, ApJ, 655, L45 
\bibitem[\protect\citeauthoryear{Mikolaitis et al.}{2017}]{2017A&A...600A..22M} Mikolaitis {\v S}., de Laverny P., Recio-Blanco A., Hill V., Worley C.~C., de Pascale M., 2017, A\&A, 600, A22 
\bibitem[\protect\citeauthoryear{Miller \& Scalo}{1979}]{1979ApJS...41..513M} Miller G.~E., Scalo J.~M., 1979, ApJS, 41, 513 
\bibitem[\protect\citeauthoryear{Moe \& Di Stefano}{2017}]{2017ApJS..230...15M} Moe M., Di Stefano R., 2017, ApJS, 230, 15 
\bibitem[\protect\citeauthoryear{Mor et al.}{2018}]{2018A&A...620A..79M} Mor R., Robin A.~C., Figueras F., Antoja T., 2018, A\&A, 620, A79 
\bibitem[\protect\citeauthoryear{Mor et al.}{2019}]{2019A&A...624L...1M} Mor R., Robin A.~C., Figueras F., Roca-F{\`a}brega S., Luri X., 2019, A\&A, 624, L1 
\bibitem[\protect\citeauthoryear{Moraux et al.}{2003}]{2003A&A...400..891M} Moraux E., Bouvier J., Stauffer J.~R., Cuillandre J.-C., 2003, A\&A, 400, 891 
\bibitem[\protect\citeauthoryear{Moraux \& Bouvier}{2012}]{2012ASSP...29..115M} Moraux E., Bouvier J., 2012, ASSP, 29, 115 
\bibitem[\protect\citeauthoryear{Olivares et al.}{2018}]{2018A&A...617A..15O} Olivares J., et al., 2018, A\&A, 617, A15 
\bibitem[\protect\citeauthoryear{Pan, Shao, \& Kulkarni}{2004}]{2004Natur.427..326P} Pan X., Shao M., Kulkarni S.~R., 2004, Natur, 427, 326 
\bibitem[\protect\citeauthoryear{Parravano, Hollenbach, \& McKee}{2018}]{2018MNRAS.480.2449P} Parravano A., Hollenbach D., McKee C.~F., 2018, MNRAS, 480, 2449 
\bibitem[\protect\citeauthoryear{Piotto \& Zoccali}{1999}]{1999A&A...345..485P} Piotto G., Zoccali M., 1999, A\&A, 345, 485
\bibitem[\protect\citeauthoryear{Portegies Zwart, McMillan, \& Gieles}{2010}]{2010ARA&A..48..431P} Portegies Zwart S.~F., McMillan S.~L.~W., Gieles M., 2010, ARA\&A, 48, 431 
\bibitem[\protect\citeauthoryear{Portinari et al.}{2004}]{2004ApJ...604..579P} Portinari L., Moretti A., Chiosi C., Sommer-Larsen J., 2004, ApJ, 604, 579 
\bibitem[\protect\citeauthoryear{Robin, Creze, \& Mohan}{1989}]{1989Ap&SS.156....9R} Robin A.~C., Creze M., Mohan V., 1989, Ap\&SS, 156, 9 
\bibitem[\protect\citeauthoryear{Robin et al.}{2003}]{2003A&A...409..523R} Robin A.~C., Reyl{\'e} C., Derri{\`e}re S., Picaud S., 2003, A\&A, 409, 523 
\bibitem[\protect\citeauthoryear{Romano, Tosi, \& Matteucci}{2006}]{2006MNRAS.365..759R} Romano D., Tosi M., Matteucci F., 2006, MNRAS, 365, 759 
\bibitem[\protect\citeauthoryear{Rybizki \& Just}{2015}]{2015MNRAS.447.3880R} Rybizki J., Just A., 2015, MNRAS, 447, 3880 
\bibitem[\protect\citeauthoryear{Salpeter}{1955}]{1955ApJ...121..161S} Salpeter E.~E., 1955, ApJ, 121, 161 
\bibitem[\protect\citeauthoryear{Scalo}{1998}]{1998ASPC..142..201S} Scalo J., 1998, ASPC, 142, 201 
\bibitem[\protect\citeauthoryear{Schroeder}{1998}]{1998A&A...334..901S} Schroeder K.-P., 1998, A\&A, 334, 901 
\bibitem[\protect\citeauthoryear{Shin \& Kim}{2016}]{2016MNRAS.460.1854S} Shin J., Kim S.~S., 2016, MNRAS, 460, 1854 
\bibitem[\protect\citeauthoryear{Silk}{1977}]{1977ApJ...214..718S} Silk J., 1977, ApJ, 214, 718 
\bibitem[\protect\citeauthoryear{Simon et al.}{2015}]{2015ApJ...808...95S} Simon J.~D., et al., 2015, ApJ, 808, 95 
\bibitem[\protect\citeauthoryear{Skrutskie et al.}{2006}]{2006AJ....131.1163S} Skrutskie M.~F., et al., 2006, AJ, 131, 1163 
\bibitem[\protect\citeauthoryear{Soderblom et al.}{2009}]{2009AJ....138.1292S} Soderblom D.~R., Laskar T., Valenti J.~A., Stauffer J.~R., Rebull L.~M., 2009, AJ, 138, 1292 
\bibitem[\protect\citeauthoryear{Sollima \& Baumgardt}{2017}]{2017MNRAS.471.3668S} Sollima A., Baumgardt H., 2017, MNRAS, 471, 3668 
\bibitem[\protect\citeauthoryear{Sollima, Cacciari, \& Valenti}{2006}]{2006MNRAS.372.1675S} Sollima A., Cacciari C., Valenti E., 2006, MNRAS, 372, 1675 
\bibitem[\protect\citeauthoryear{Sollima, Ferraro, \& Bellazzini}{2007}]{2007MNRAS.381.1575S} Sollima A., Ferraro F.~R., Bellazzini M., 2007, MNRAS, 381, 1575 
\bibitem[\protect\citeauthoryear{Spitzer}{1940}]{1940MNRAS.100..396S} Spitzer L., Jr., 1940, MNRAS, 100, 396 
\bibitem[\protect\citeauthoryear{Spitzer}{1987}]{1987degc.book.....S} Spitzer L., 1987, in "Dynamical evolution of globular clusters", Princeton, NJ, Princeton University Press
\bibitem[\protect\citeauthoryear{Tognelli, Prada Moroni, \& Degl'Innocenti}{2011}]{2011A&A...533A.109T} Tognelli E., Prada Moroni P.~G., Degl'Innocenti S., 2011, A\&A, 533, A109 
\bibitem[\protect\citeauthoryear{Udalski, Szymanski, Soszynski \& Poleski}{2008}]{2008AcA....58...69U} Udalski A., Szymanski M.~K., Soszynski I., Poleski R., 2008, AcA, 58, 69
\bibitem[\protect\citeauthoryear{van Dokkum \& Conroy}{2010}]{2010Natur.468..940V} van Dokkum P.~G., Conroy C., 2010, Natur, 468, 940 
\bibitem[\protect\citeauthoryear{Wang et al.}{2019}]{2019MNRAS.482.2189W} Wang C., et al., 2019, MNRAS, 482, 2189 
\bibitem[\protect\citeauthoryear{Weisz et al.}{2013}]{2013ApJ...762..123W} Weisz D.~R., et al., 2013, ApJ, 762, 123 
\bibitem[\protect\citeauthoryear{Zakharova}{1989}]{1989AN....310..127Z} Zakharova P.~E., 1989, AN, 310, 127 


\end{thebibliography}
\end{document}